\documentclass[11pt]{article}
\usepackage[textwidth=15.2cm,textheight=22cm]{geometry}

\def\XXX#1{}
\def\kpm{}
\def\kmp{-}
\def\Chk#1{}
\def\Rder#1{}

\def\fracpp#1#2{\frac{\partial #1}{\partial #2}}

\def\non{\nonumber}

\def\tA{\totalarea}
\def\totalarea{[\sigma]}

\def\Tr{\mbox{Tr}}

\def\tensorp{\otimes}

\def\mC{\mathcal{C}}
\def\mN{\mathcal{N}}
\def\mL{\mathcal{L}}
\def\mP{\mathcal{P}}

\def\a{\alpha}
\def\b{\beta}
\def\c{\gamma}
\def\d{\delta}
\def\e{\epsilon}
\def\k{\kappa}
\def\m{\mu}
\def\n{\nu}
\def\r{\rho}
\def\s{\sigma}

\def\t{\tau}

\def\O{\Omega}
\def\p{\pi}
\def\P{\Pi}
\def\th{\theta}
\def\C{\Gamma}

\def\bz{\bar{z}}
\def\bw{\bar{w}}
\def\bI{\bar{I}}
\def\bJ{\bar{J}}

\def\halpha{{\hat{\alpha}}}

\def\hmu{{\hat{\mu}}}
\def\hnu{{\hat{\nu}}}
\def\hrho{{\hat{\rho}}}
\def\hsigma{{\hat{\sigma}}}

\def\hz{{\hat{z}}}
\def\hw{{\hat{w}}}
\def\ha{{\hat{a}}}
\def\hb{{\hat{b}}}
\def\hc{{\hat{c}}}
\def\hd{{\hat{d}}}

\def\hA{{\hat{A}}}
\def\hB{{\hat{B}}}
\def\hC{{\hat{C}}}
\def\hD{{\hat{D}}}
\def\hE{{\hat{E}}}
\def\hF{{\hat{F}}}
\def\hx{{\hat{x}}}

\def\PLB#1#2#3{Phys. Lett. {\bf B#1}(#2)#3}

\begin{document}
\begin{titlepage}
\begin{flushright}    
{\small AEI-2008-025}

\end{flushright}
\vskip 1cm

\centerline{\LARGE{\bf { $\beta$-deformation for matrix model of M-theory }}}
\vskip .3cm

\vskip 1.5cm

\vskip 1cm

\centerline{
Hidehiko Shimada
}

\vskip 1cm

\centerline{{\it {Max-Planck-Institut f\"{u}r 
Gravitationsphysik}}}
\centerline{{\it {$\;$ Albert-Einstein-Institut, Potsdam, Germany}}}

\vskip 0.5cm

\vskip 1.5cm

\centerline{\bf {Abstract}}
\vskip .5cm

\noindent 
A new class of deformation of the matrix model of M-theory
is considered. 
The deformation is 
analogous to the so-called $\b$-deformation
of $D=3+1$, $\mN=4$ Super Yang-Mills theory,
which preserves the conformal symmetry.
It is shown that the deformed matrix
model can be considered as a matrix model
of M-theory on a certain curved background in eleven-dimensional supergravity,
under a scaling limit involving 
the deformation parameter and $N$ (the size of the matrices). 
The background belongs to the so-called pp-wave type metric
with a non-constant four-form flux depending 
linearly on transverse coordinates.
Some stable solutions of the deformed model
are studied, which correspond to membranes 
with the torus topology.
In particular, it is found that apparently distinct configurations of membranes,
having different winding numbers, are indistinguishable in the matrix model.
Simultaneous introduction of both $\b$-deformation and mass-deformation
is also considered, and, in particular, a situation is found
in which the stable membrane configuration interpolates
between a torus and a sphere, depending on the values of the deformation parameters.

\vfill

\end{titlepage}


\section{Introduction}
\label{RSIntro}
Although M-theory~\cite{RBMTheoryHT}\cite{RBMTheoryW} plays a crucial role in
non-perturbative physics of String theory,
its formulation is not yet established.
The best candidate so far,
the matrix model of M-theory~\cite{RBdeWitHoppeNicolai}\cite{RBBFSS}, 
has fundamental unsolved
problems such as the problem of $N \rightarrow \infty $
limit and the eleven-dimensional Lorentz invariance.
Another important issue is the relation of the matrix model
to the supergravity background.
As the matrix model should contain degrees of freedom of eleven-dimensional
supergravity, condensation of them 
should in principle yield the matrix model on curved backgrounds.
Also, the information of the supergravity equations of motion 
should somehow be incorporated into the matrix model
formulation of M-theory.

An attractive approach to these problems is to consider
the matrix model as a 
regularised version~\cite{RBMR}\cite{RBdeWitHoppeNicolai} 
of supermembrane
theory~\cite{RBBergshoeffSezginTownsendShort}\cite{RBBergshoeffSezginTownsendLong};
the large $N$ limit can be interpreted as the renormalisation
of membrane theory, and the Lorentz generators are known in 
membrane theory~\cite{RBLorentzMembrane}. 
Also, the relation of supermembrane theory to
the background equation of motion is well 
understood~\cite{RBBergshoeffSezginTownsendShort}\cite{RBBergshoeffSezginTownsendLong}.

However, we are still far from the complete resolution to these issues,
and it is necessary to gain more experience of and
insight into the physics of
membranes and the matrix model.
In this paper, we will consider a new deformation of the matrix model,
based on an analogy to four-dimensional gauge theory.
The model rather unexpectedly turns out to be equivalent
to a regularised membrane theory on a certain curved background.
General motivation to study this deformation would be twofold.
First, it will be useful to have explicit examples,
in order to understand the general relation
between the matrix model and backgrounds.
Second, as the deformed model has parameters which can be controlled
freely, one might expect to find
tractable and interesting physics by tuning them.
Indeed, we find that the deformed model
has stable solutions, which 
correspond to membranes with torus topology.
\XXX{*** for some particular values of the deformation parameter}

The explicit form of our deformation
is motivated by the following consideration.
The matrix model of M-theory 
and four-dimensional $\mN=4$ supersymmetric Yang-Mills theory (SYM)
are similar in many ways;
in particular, they both have the maximal supersymmetry
which is highly restrictive.
\XXX{
which severely restricts the action and dynamical properties
of them.}
The $\mN=4$ SYM has conformal symmetry as well,
and gives a prime example of a fixed line
of the renormalisation group flow in the theory space
of four-dimensional field theory.
The deformations of $\mN=4$ SYM which preserve the conformal symmetry
are interesting from this point of view, and have been studied extensively, 
in particular for the case where the $\mN=1$ supersymmetry is also preserved.
One class of such deformations
is the $\beta$-deformation with single deformation 
parameter~\cite{RBBetaDeformationEarly}\cite{RBBetaDeformationPiguetEtal}\cite{RBBetaDeformationLeighStrassler}.
Recently this deformation was revisited in the context of 
the AdS/CFT correspondence~\cite{RBLuninMaldacena}, and
was generalised to a deformation with three parameters where the supersymmetry is
in general completely broken~\cite{RBFrolov}.
For field-theoretic discussion and proofs of the conformal invariance (or 
the scale invariance) of the $\b$-deformed theory in general, 
see
\cite{RBBetaDeformationEarly}\cite{RBBetaDeformationRecent}\cite{RBBetaDeformationLeighStrassler}\cite{RBBetaDeformationPiguetEtal}\cite{RBBetaDeformationAKS}.

The $\beta$-deformation (including its non-supersymmetric generalisation)
consists in modifying the Yukawa couplings and 
the quartic couplings of scalar fields by
certain phase factors.~\footnote{
Actually, this prescription is only true 
in the leading order of $1/N$.
In order to maintain the scale invariance,
one in general needs to introduce $1/N$
corrections to various couplings.
We 
will comment on this issue for the matrix model 
at the end of section \ref{RSSolution}.
}
As the matrix model of M-theory has similar Yukawa and
scalar quartic couplings,
phase factors can be introduced in a similar 
manner.
It therefore seems natural to study this deformation of the matrix model,
and consider whether it has also some significance.

One of the main results of this paper is that
this deformed matrix model, introduced from a rather mathematical
analogy to four-dimensional theory,
indeed admits an interpretation from the M-theory point of view.
We shall show that this model, under a certain scaling limit involving
both $N$ (the size of matrices)
and the deformation parameter, can be considered as a matrix model of 
M-theory on a certain curved background,
and that the background solves the supergravity equations of motion.
We do this by showing that the matrix model arises
from regularisation of supermembrane
theory on that background.
The background belongs to
the so-called pp-wave (or plane-wave) backgrounds
and is supported by a {\it non-constant}
four-form flux.
The pp-wave background with a non-constant flux
in eleven-dimensional supergravity is 
first considered in \cite{RBPPwaveNonconstant}.

In \cite{RBBMN}, another deformed matrix model,
the BMN matrix model, was proposed,
which is characterised by mass terms for scalars and fermions,
and by cubic scalar couplings. In \cite{RBBMNMembrane1}\cite{RBBMNMembrane2}
it was shown that
this model is equivalent to regularised supermembrane 
theory on a supergravity background,
which is  also of the pp-wave type, but is supported
by a constant four-form flux.
This analysis 
for the original maximally supersymmetric BMN matrix model
was later generalised to less supersymmetric models~\cite{RBIizuka}.
We also mention that a matrix model similar to ours are considered
and used in \cite{RBBerensteinMMAdSCFT} to understand the AdS/CFT correspondence.

The organisation of this paper is as follows.
We first define our deformed matrix model in section \ref{RSDeformation}.
Its supersymmetry is also considered.
Section \ref{RSBackground} is devoted to
establishing the equivalence between the deformed matrix model
and the regularised supermembrane theory on the background.
\XXX{the background?}
In section \ref{RSSolution} we discuss some stable solutions of the model.
The solutions correspond to membranes with the topology of a torus 
wrapped in general several times on a certain $S^1\times S^1$.
We show that some physically distinct configurations
in conventional membrane theory
are indistinguishable in the matrix model.
In section \ref{RSMass} we consider the matrix model
associated with the background which involve both 
our deformation parameters and the BMN-like mass parameters.
In particular, we find a class of models where the
stable membrane configuration has the topology of either a torus or a sphere,
depending on the values of the deformation parameters.
We conclude in section \ref{RSConclusion} with some discussion.

\section{Deformation} 
\label{RSDeformation}
In this section we describe the deformation,
which is motivated by an analogy to the $\beta$-deformation
(and its non-supersymmetric generalisation)
of $\mathcal{N}=4$ SYM in four dimension.
The deformation can be described succinctly by using the
$*$-product notation~\cite{RBLuninMaldacena}\cite{RBFrolov}
explained below.

The original matrix model is governed by the Hamiltonian
\begin{equation}
H=\Tr \left( \frac{1}{2} (\P^\alpha)^2 
- \frac{1}{4} [X^\alpha, X^\beta]^2
+ \Psi^T \c^\a [X_\a,\Psi]\right),
\label{RFHamiltonianOriginalMatrixModel}
\end{equation}
\XXX{SIGN BEFORE YUKAWA, MANIPULATED TO 
MEMBRANE WITHOUT XNEW=-XOLD}
and the phase space constraints corresponding to the $U(N)$ gauge symmetry,
\begin{equation}
[X^\a,\P^\a]
-2 i \Psi^T \Psi
=0, \label{RFConstraintUN}
\end{equation}
where $X^\a$, $(\a=1,\dots, 9)$, are $N\times N$ hermitian matrices and
$\Pi^\a$ are their conjugate momenta, and 
$\Psi^a$, $(a=1,\dots,16)$, are fermionic $N \times N$ hermitian matrices
which are canonically conjugate to themselves. The relevant Dirac brackets are,
\begin{eqnarray}
\{
{(X^\a)^i}_j,{(\Pi^\b)^k}_l
\}_{\mbox{D.B.}}
&=&
\delta^{\a\b} {\d^i}_l {\d^k}_j, \qquad (i,j,k,l =1,\cdots,N),
\label{RFDBOriginalMMBoson}
\\
\{{(\Psi^a)^i}_j,{(\Psi^b)^k}_l\}_{\mbox{D.B.}}
&=&
-\frac{i}{2}
\delta^{ab} {\d^i}_l {\d^k}_j, \qquad (i,j,k,l =1,\cdots,N).
\label{RFDBOriginalMMFermion}
\end{eqnarray}
The model has a SO(9) symmetry under which 
$X$ transforms as a SO(9) vector and $\Psi$ as a 16-component
real spinor.  We choose a real and symmetric representation of
$16\times 16$ gamma matrices $\gamma^\alpha$.
Equivalently, the model is described by the action
\begin{equation}
S=
\int
\Tr
\left(
\frac{1}{2}\left(D_0 X^\alpha\right)^2
+
\frac{1}{4}
[X^\a, X^\b]^2
+i \Psi^T D_0 \Psi
-\Psi^T \c^\a [X^\a,\Psi]
\right)
dt,
\label{RFActionOriginalMatrixModel}
\end{equation}
\XXX{SIGN BEFORE YUKAWA, MANIPULATED}
where the covariant derivative is given by
$D_0 f= \partial_0 f - [-i A_0, f]$.

%

The class of deformation we consider is in general parametrised 
by six parameters.
Before describing the general deformation
we will focus on a particular case
which is parametrised by a single-parameter $\b$
as it is much easier to grasp.

We should first introduce some notations.
We choose
two commuting U(1) charges
in the ``flavour'' SO(9) symmetry, 
the rotation in the 12 plane and 34 plane, 
and call them as $Q_{(1)}$ and $Q_{(2)}$.
We define complex combinations of scalars,
\begin{equation}
Z= \frac{X^1+ iX^2}{\sqrt{2}}, \quad W=\frac{X^3+ i X^4}{\sqrt{2}},
\end{equation}
which have definite $Q_{(1)}, Q_{(2)}$ charges.
We denote the $U(1)$ charges of a field $f$ appearing in the matrix model Hamiltonian
by $Q^f_{(1)}$ and $Q^f_{(2)}$; for example, $Q^Z_{(1)}=1$,
$Q^Z_{(2)}=0$ and $Q_{(2)}^{W^\dagger}=-1$.

We then introduce the $*$-product by
\begin{equation}
f*g=e^{i \pi \b(
Q^f_{(1)}
Q^g_{(2)}
-
Q^g_{(1)}
Q^f_{(2)})
} f g.
\end{equation}
In this paper we will only consider the case where $\b$ is real.
Thus the $*$-product is the usual product simply modified
by a flavour-dependent phase factor. 

Our deformation consists in 
replacing all 
commutators appearing in the original matrix model 
Hamiltonian (\ref{RFHamiltonianOriginalMatrixModel}), or, 
equivalently, in the action
(\ref{RFActionOriginalMatrixModel}),
by the $*$-commutator defined by
\begin{equation}
[f,g]_*=f * g - g * f.
\end{equation}
Thus, the Hamiltonian and the action of the deformed model are,
\begin{eqnarray}
H&=&\Tr \left( \frac{1}{2} (\P^\alpha)^2 
- \frac{1}{4} ([X^\alpha, X^\beta]_*)^2
+ \Psi^T \c^\a [X_\a,\Psi]_*\right),
\label{RFHamiltonianDeformedMM}
\\
S&=&
\int
\Tr
\left(
\frac{1}{2}\left(D_0 X^\alpha\right)^2
+
\frac{1}{4}
\Bigg([X^\a, X^\b]_*\right)^2
+i \Psi^T D_0 \Psi
-\Psi^T \c^\a [X^\a,\Psi]_*
\Bigg)
dt.
\label{RFActionDeformedMM}
\end{eqnarray}
\XXX{SIGN BEFORE YUKAWA,MANIPULATED}
\XXX{Firstclass?}
The phase space constraints (\ref{RFConstraintUN}) are unchanged.
In the above formulae, relevant expressions in the bosonic potential term 
are, 
\begin{eqnarray}
[Z, W]_{*}&=& e^{i \p \b} Z W - e^{-i \p \b } W Z, \\\nonumber
[Z, W^\dagger]_{*}&=& e^{-i \p \b} Z W^\dagger - e^{+i \p \b } W^\dagger Z,
\end{eqnarray}
and its complex conjugate.
Other commutators such as $[X^5, Z]$ or $[Z, Z^\dagger]$ are left unchanged.
For fermionic terms, 
we need projectors
such as 
$(1\pm \c^{\bz z})/2$ which pick up 
components with $Q^{(1)}$-charge $\pm 1/2$.\footnote{
We use $\c^{z\bz}=\frac{1}{2}(\c^z \c^{\bz} - \c^{\bz} \c^z)$,
$\c^{z}=\frac{1}{\sqrt{2}}(\c^1+i\c^2)$
and $\c^{\bz}=\frac{1}{\sqrt{2}}(\c^1-i\c^2)$.
}
For example, we have
\begin{eqnarray}
Z*\Psi
&=&Z*\left(
\frac{1+\c^{\bw w}}{2}\Psi
+
\frac{1-\c^{\bw w}}{2}\Psi
\right)
\non
\\
&=&
e^{\frac{i \pi \b}{2}}
Z
\frac{1+\c^{\bw w}}{2}\Psi
+
e^{-\frac{i \pi \b}{2}}
Z
\frac{1-\c^{\bw w}}{2}\Psi,
\label{RFStarcommutatorExampleScalarFermion}
\end{eqnarray}
which can also be written as
\begin{equation}
Z*\Psi
=
Z
e^{i\pi\b\frac{1}{2}\c^{\bw w}}
\Psi.
\end{equation}

The generalisation of this one-parameter deformation 
is obtained just by extending the definition of the $*$-product
to include more general $U(1)$ generators in the $SO(9)$ symmetry.
There are four independent commuting $U(1)$ charges. We choose 
the rotations in the 12, 34, 56, 78 planes,
and label them by indices $I, J, \cdots= 1, 2, 3, 4$.
For each pair of $U(1)$ charges $(I, J)$ a deformation parameter,
$\beta^{(I J)}=-\beta^{(J I)}$, can be introduced.
Hence there are six independent parameters.
The generalised $*$-product is now given by
\begin{equation}
f*g=
\left( e^{i \p \sum_{I<J}\b^{(I J)} 
\left(Q^f_{(I)} Q^g_{(J)}
-Q^f_{(J)} Q^g_{(I)}\right)}
\right)f g.
\end{equation}
The one-parameter deformation described before is the special case where
the only non-zero deformation parameter is $\b^{(1 2)}=\b(=-\b^{(2 1)})$.

We group eight hermitian scalar fields into
four complex fields $Z^I, (I=1,2,3,4)$ as
\begin{eqnarray}
Z^1&=&\frac{1}{\sqrt{2}}(X^1+iX^2),\\
Z^2&=&\frac{1}{\sqrt{2}}(X^3+iX^4),\\
Z^3&=&\frac{1}{\sqrt{2}}(X^5+iX^6),\\
Z^4&=&\frac{1}{\sqrt{2}}(X^7+iX^8).
\end{eqnarray}
The generalised $*$-commutators for bosonic fields 
in (\ref{RFHamiltonianDeformedMM}) and (\ref{RFActionDeformedMM}) are
now given by,
\begin{equation}
[Z^I, Z^J]_{*}=
e^{i\p\b^{(IJ)}} Z^I Z^J
-
e^{-i\p\b^{(IJ)}} Z^J Z^I
\end{equation}
and
\begin{equation}
[Z^I,Z^{\dagger J}]_{*}=
e^{-i\p\b^{(IJ)}} Z^I Z^{\dagger J}
-
e^{i\p\b^{(IJ)}} Z^{\dagger J} Z^I.
\end{equation}
For indices $I,J$ we will not imply the summation over repeated indices.

In general the $SO(9)$ symmetry is broken down into 
the $U(1)^4$ symmetry spanned by $Q_{(I)}$'s.
If $\b^{(I 4)}=0$, the matrix model
can be considered as a result of dimensional reduction of 
the three-parameter deformation of $D=3+1$, $\mN=4$ SYM
introduced in~\cite{RBFrolov}.
Furthermore if 
$
\b^{(1 2)}
=
\b^{(2 3)}
=
\b^{(3 1)}
$ the model is a dimensionally reduced
form of the $D=3+1$, $\beta$-deformed $\mN=4$ SYM
with $\mN=1$ supersymmetry,
and has corresponding supersymmetry.
\XXX{THINK FROM SUGRA POINT OF VIEW}

The deformation in general breaks both kinematical and dynamical supersymmetry of 
the original matrix model. 
However, for special values of the deformation parameters,
a part of the supersymmetry remain unbroken, provided
that the supersymmetry transformation law is appropriately 
modified.
For these special values,
there exist 
16-component spinors $\delta \xi$
whose U(1)-charges $s_I=\pm\frac{1}{2}$ satisfy
\begin{equation}
\sum_J 
\b^{(IJ)}
s_J
=0,
\label{RFStarNeutral}
\end{equation}
for all $I=1,2,3,4$.
This relation is equivalent to the condition
that the $*$-product between $\delta \xi$ 
and any field reduces to the ordinary product.
This property of $\d \xi$, which we call the $*$-neutrality,
ensure the invariance of the action (\ref{RFActionDeformedMM})
under the modified supersymmetry transformation 
with $\d \xi$ as the infinitesimal parameter.
This is true for both dynamical and kinematical supersymmetry.
The modified transformation law for the dynamical supersymmetry is
given by,
\begin{eqnarray}
&&
\d X^\a = i \Psi^T \c^\a \d \xi,
\\
&&\d \Psi= \left(
\frac{1}{2} D_0 X^\a \c^\a
-\frac{i}{4}
[X^\a, X^\b]_* \c^{\a\b} \right)
\d \xi.
\end{eqnarray}
\Chk{*}
Because of the $*$-neutrality,
one can move $\d \xi$ in the variation
of the action, without producing extra phase factors.
Then one can show the invariance of the action,
in the same manner as in the original matrix model,
using the associativity of the $*$-product 
and the property that if the product $fg$ is uncharged, $f*g=fg$.
The transformation law for the kinematical supersymmetry 
is not modified and given by,
\begin{equation}
\d \Psi= \d \xi {\bf 1}, \quad \d X^\a =0,
\end{equation}
where ${\bf 1}$ is the $N\times N$ unit matrix.

Let us investigate the condition on parameters $\b^{(IJ)}$ 
under which 
$*$-neutral spinors satisfying (\ref{RFStarNeutral}) 
exist.
The 16-component spinors 
can be spanned by the following basis vectors,
\begin{equation}
|s_1,s_2,s_3,s_4\rangle,
\label{RFBasis16spinor}
\end{equation}
labelled by 
the eigenvalues of the four U(1) charges.
We abbreviate, for instance, 
$|\frac{1}{2},\frac{1}{2},-\frac{1}{2},\frac{1}{2}\rangle$ by
$|+,+,-,+\rangle$.

It is sufficient to consider the case where 
$|+,+,+,+\rangle$ is $*$-neutral, which
automatically implies that 
$|-,-,-,-\rangle$ is also $*$-neutral.
Choosing other element of the basis
(\ref{RFBasis16spinor}) is related 
by a simple redefinition of $\b$'s. For example,
using $|+,+,+,-\rangle$ instead of  $|+,+,+,+\rangle$ amounts to
flipping the sign of $\b^{(I 4)},\  I=1,2,3$.
From (\ref{RFStarNeutral}), one finds four linear equations,
for six variables $\b^{(I J)}$.
Actually, it is easy to see that only three of the equations 
are linearly independent,
and hence $\b$'s are parametrised by three parameters.
Concretely, we choose $\b^{(1 2)}$, $\b^{(2 3)}$, $\b^{(3 1)}$,
and express the others by,
\begin{equation}
\b^{(1 4)} = \b^{(3 1)} - \b^{(1 2)}, 
\ 
\b^{(2 4)} = \b^{(1 2)} - \b^{(2 3)}, 
\ 
\b^{(3 4)} = \b^{(2 3)} - \b^{(3 1)}. 
\end{equation}
\Chk{*}
Under this condition, the deformed model
has the dynamical and kinematical supersymmetry,
each with two-component supercharges.

To consider the case of the higher supersymmetry,
there are two essentially distinct possibilities,
namely, to add 
(a) $|+,+,+,-\rangle$ and $|-,-,-,+\rangle$,
or 
(b)
$|+,+,-,-\rangle$ and $|-,-,+,+\rangle$, as
$*$-neutral spinors.
After reducing eight linear equations following from
(\ref{RFStarNeutral}) to independent equations,
one finds that the the possibility (a) leads to 
the single-parameter deformation with
\begin{equation}
\b^{(I 4)}=0,\quad \b^{(1 2)} = \b^{(2 3)} = \b^{(3 1)},
\label{RFStarNeutralFour1}
\end{equation}
which is equivalent to the condition that the deformed model
is the dimensionally reduced version of $D=3+1$, $\b$-deformed SYM
with the $\mN=1$ supersymmetry, discussed before.
The case (b) yields the condition
\begin{equation}
\b^{(12)}=0,\quad
\b^{(34)}=0,\quad
\b^{(14)}=
\b^{(31)}=
\b^{(23)}=
\b^{(24)}, \label{RFStarNeutralFour2}
\end{equation}
\Chk{*}
which also give a single-parameter deformation.
These conditions 
(\ref{RFStarNeutralFour2}) cannot be made 
equivalent to (\ref{RFStarNeutralFour1}) by 
reshuffling of the coordinates; one can show that the bosonic flavour
symmetry in this case is completely broken down into $U(1)^4$ symmetry,
whereas in the case of (\ref{RFStarNeutralFour1}), 
the flavour SO(3) symmetry (the rotation within $X^7, X^8, X^9$ directions)
is present. 
If we further increase the 
number of $*$-neutral spinors,
we only arrive at the trivial case where all $\b$'s are zero.
We have thus essentially completed the classification
of the supersymmetry of the $\b$-deformed matrix model.

\XXX{Algebra is simple}

\section{Deformed model and D=11 SUGRA background}
\label{RSBackground}
The aim of this section is to show that the deformed matrix model,
described in section \ref{RSDeformation},
in a certain scaling limit,
is equivalent 
to a regularised membrane theory on a certain curved
background of eleven-dimensional supergravity.
This makes a good case for considering 
the deformed model as the matrix model of
M-theory on that background. 

We will begin by a brief review, in subsection \ref{RSSReviewFlatspaceMR}, of
bosonic membrane theory on flat spacetime in the lightcone gauge,
and the regularisation procedure, the matrix regularisation, 
in order to collect necessary formulae.
Then we describe, in \ref{RSSScalingLimit},
how the scaling limit naturally arises
from consideration of the $*$-commutator in the light of
matrix regularisation.
Under this scaling limit, we then describe the continuum theory
corresponding to the deformed model, in \ref{RSSBackground}, and 
show that this continuum theory is equivalent to lightcone membrane theory
on a certain background. We then show that this background solves
the equations of motion of eleven-dimensional supergravity,
including an overall factor.
Up to this point, we will confine ourselves to the bosonic degrees of freedom of 
membrane theory. In \ref{RSSFermionic}, we analyse  
the fermionic degrees of freedom, and show that the 
fermionic sector of the deformed model
precisely matches with that of regularised
lightcone supermembrane theory on the background.

\subsection{Review of bosonic membrane in lightcone gauge and matrix regularisation
on flat spacetime}
\label{RSSReviewFlatspaceMR}
This subsection consists of a brief review
of the
lightcone gauge formalism (in the spirit of  \cite{RBGGRT})
for the bosonic part of membrane theory on flat spacetime, 
and the matrix regularisation procedure which turns lightcone membrane theory
into the matrix model.

The action of the bosonic part of membrane theory on 
flat spacetime is given by its 3-volume,
\begin{eqnarray}
&&S= \int \mL d^2\s d\t =-T \int \sqrt{-\det (h_{ij})} \  d^2\s d\t,\\ 
&&h_{i j}= \eta_{\m \n} \fracpp{x^\m}{\s^i}\fracpp{x^\n}{\s^j}, \qquad
(i,j=0,1,2), \non
\label{RFActionMembraneFlatBosonic}
\end{eqnarray}
where $x^\mu(\s^0, \s^1, \s^2)=
x^\mu(\t, \s^1, \s^2)$, $(\m= 0,1, \cdots, 10)$,
gives the parametrisation of the membrane worldvolume
embedded in spacetime, 
and $T$ is the membrane tension. We hereafter mostly work in the length scale in which
$T=1$. 
The canonical momenta 
\begin{equation}
\mP_\mu = \fracpp{\mL}{\left(\partial_\t x^\mu\right)} 
\end{equation}
satisfy the following identities
(phase space constraints), which represent the
reparametrisation invariance of the action,
\begin{eqnarray}
&&\eta^{\m \n} \mP_\mu \mP_\nu + \frac{1}{2} \{x^\mu, x^\nu\}^2 = 0, 
\label{RFConstraintTau}
\\
&&\mP_\mu \fracpp{x^\mu}{\s^r} =0, \qquad (r=1, 2), \label{RFConstraintSigma}
\end{eqnarray}
where we have defined
\begin{equation}
\{f, g\}= \fracpp{f}{\s^1}\fracpp{g}{\s^2} - \fracpp{f}{\s^2}\fracpp{g}{\s^1},
\end{equation}
where $f$ and $g$ are functions defined on the $(\s^1,\s^2)$-space.
We shall call this structure, analogous to the Poisson brackets
(for a system with one degree of freedom), as the Lie brackets in this paper.

In the lightcone gauge, we first identify the
$\t$ coordinate with the spacetime coordinate 
$x^+$,\footnote{
Our lightcone convention is $x^{\pm}=\frac{1}{\sqrt{2}}(x^0\pm x^{10})$.}
\begin{equation}
\t=x^+. \label{RFGaugeFixTau}
\end{equation}
We then partially fix $(\s^1, \s^2)$ coordinates by
requiring the momentum density $\mP_-$ to be constant
in the $(\s^1, \s^2)$ directions, or equivalently,
\begin{equation}
\mP_-= \frac{P_-}{\tA}, \label{RFGaugeFixSigma}
\end{equation}
where $P_-=\int \mP_- d^2\s$ is the total 
momentum in the $-$ direction  
and $\tA=\int d^2 \s$ is a constant 
representing the total area of the base space $(\s^1, \s^2)$.
\footnote{
The constant $\tA$ depends on conventions,
and cancels out in any relation between physical observables.} 

These gauge fixing conditions (\ref{RFGaugeFixTau})(\ref{RFGaugeFixSigma}) 
allows one to explicitly solve the
phase space constraints (\ref{RFConstraintTau})(\ref{RFConstraintSigma});
we have,
\begin{eqnarray}
&&\fracpp{x^-}{\s^r}= -\frac{\tA}{P_-}\mP^\a 
\fracpp{x^\a}{\s^r},
\qquad(r=1,2),
\label{RFBosonicSolvedXm}
\\
&&-\mP_+
=
\frac{\tA}{2(-P_-)} 
\left(
\left(\mP^\alpha\right)^2
+
\frac{1}{2}\{x^\a, x^\b\}^2
\right). \label{RFBosonicSolvedPp}
\end{eqnarray}
Here indices $\a, \b$ 
for transverse directions run through $1$ to $9$.

From (\ref{RFBosonicSolvedPp}),
we obtain the Hamiltonian 
\begin{equation}
H=-P_+ = 
\int\left(-\mP_+\right)d^2\s
=
\int
\frac{\tA}{2(-P_-)} 
\left(
\left(\mP^\alpha\right)^2
+
\frac{1}{2} \{x^\a, x^\b\}^2
\right)d^2\s.
\label{RFHamiltonianFlatBosonic}
\end{equation}
We put the factor $-1$ before the lightcone component 
of the momentum because 
$-\mP_->0$ and $-\mP_+ >0$ hold.
The equation (\ref{RFBosonicSolvedXm}) implies 
the integrability condition
\begin{equation}
\{x^\a, \mP^\a\}=0. \label{RFConstraintAPD}
\end{equation}
This equation acts as a phase space constraint of lightcone membrane theory,
and corresponds to the residual reparametrisation
invariance by area preserving diffeomorphisms
acting on the $(\s^1,\s^2)$-space.

The bosonic sector of the 
original matrix model, described by 
the Hamiltonian (\ref{RFHamiltonianOriginalMatrixModel})
and the constraint (\ref{RFConstraintUN}),
can be considered as
a regularised version of the continuum theory 
described by (\ref{RFHamiltonianFlatBosonic})
and (\ref{RFConstraintAPD}).
Let us recall basic relations involved in 
this matrix regularisation. 
In matrix regularisation,
functions $f(\s^1, \s^2), g(\s^1, \s^2), \cdots$ 
are turned into $N\times N$ matrices $\hat{f}=\rho(f), \hat{g}=\rho(g), \cdots$. 
These matrices give discrete approximation to 
the corresponding functions.
Some operations acting on functions have 
counterparts acting on the corresponding matrices.
This correspondence, apart from the linearity of $\rho(f)$,
can be summarised as follows,
\begin{eqnarray}
\rho(f g) &\approx& \frac{1}{2} \left(\rho(f) \rho(g) + \rho(g) \rho(f)\right),
\label{RFMRMultiplication}
\\
\rho\left(\{f, g\}\right) 
&\approx& -i \frac{2\pi N}{\tA} \left[ \rho(f), \rho(g) \right],
\label{RFMRCommutator}
\\
\frac{1}{\tA}\int f d^2 \s &\approx& \frac{1}{N} \Tr \rho(f).
\label{RFMRTr}
\end{eqnarray}
Left-hand sides and right-hand sides of these formulae
are equal up to higher order corrections in $1/N$.
The first relation means that
multiplication of two functions 
corresponds to multiplication (more precisely 
taking one-half of the anti-commutator) of the corresponding
matrices.
The second relation then tells us that Lie brackets
between two functions correspond to the commutator of the corresponding matrices
multiplied by a factor proportional to $N$. 
\footnote{
The factor before the commutator in (\ref{RFMRCommutator})
can be understood as follows.
By using the well-known
mathematical analogy between matrix regularisation and quantisation of 
a system with single degree of freedom,
it corresponds to $1/(i \hbar)$ in quantum mechanics. 
Now, every state vector in quantum mechanics 
occupies the area $2\pi \hbar$ in 
the $(x,p)$-space (in the semi-classical regime).
In matrix regularisation there are $N$ independent ``state vectors'',
and the $(\s^1,\s^2)$-space is divided into $N$ parts with equal area 
$\tA/N$. Hence $\tA/N$ corresponds to $2\pi \hbar$, and
 $-i 2\pi N/\tA$ to $1/(i\hbar)$.
}
In particular, this relation implies 
that the commutator of two matrices of order unity is of order $1/N$. 

After an appropriate rescaling of the matrices and the time coordinate, 
the matrix-regularised Hamiltonian 
(and the constraint)
becomes identical to that of the matrix model.
See appendix \ref{RSARescaling} for details.

\subsection{$*$-commutator and scaling limit}
\label{RSSScalingLimit}
The first step towards the continuum version of
the deformed matrix model 
is to find the continuum counterpart of the $*$-commutator. 
A scaling limit involving $N$ and the deformation parameters
$\b$ naturally arises in this consideration.
This scaling limit plays an essential role in this paper.

We first focus on the single-parameter deformation.
Defining $\hat{z}=\rho(z), \hat{w}=\rho(w)$, we have
\begin{eqnarray}
[\hz, \hw]_{*}&=& e^{i \p \b} \hz \hw - e^{-i \p \b } \hw \hz
\nonumber\\
&\approx& [\hz, \hw] + i 2\pi \b \frac{1}{2} (\hz \hw + \hw \hz),
\label{RFStarCommutatorExpanded}
\end{eqnarray}
where we have assumed $\beta \ll1$. The constant rescaling 
between $\hz, \hw$ and $Z, W$, 
noted at the end of \ref{RSSReviewFlatspaceMR},
is without effect, as all expressions in this subsection are homogeneous in
$\hz$ and $\hw$.

As noted below (\ref{RFMRCommutator}), the commutator 
term above is of order $1/N$. Hence in the regime where $\b$ is
also of order $1/N$, or equivalently, if we fix $\b N$ when taking $N$ large,
the two terms in (\ref{RFStarCommutatorExpanded}) are comparable 
and both contribute to the dynamics of membranes.
Throughout this paper, we shall assume this scaling limit.~\footnote{
The decomposition of the $*$-commutator into a commutator
and an anti-commutator piece is possible even if $\b \sim 1$. 
In this case, the commutator term, 
which represents the effect of the membrane tension, becomes negligible
compared to the anti-commutator term. 
This limiting case, which might be called
as the ``membrane bit'' regime, might also be interesting.
}
We stress that
the deformation remains non-trivial in the $N \rightarrow \infty$ limit,
albeit the scaling limit $\b \sim 1/N$, since the commutator term
already is of order $1/N$.
Then, applying (\ref{RFMRCommutator}) and (\ref{RFMRMultiplication})
to (\ref{RFStarCommutatorExpanded}), we obtain
\begin{eqnarray}
[\hz, \hw]_{*}
&\approx& 
i \frac{\tA}{2 \pi N} \left(
\rho(\{z, w\}) + \b N \frac{(2\pi)^2}{\tA} \rho( z w )
\right)\\
&=&
i \frac{\tA}{2 \pi N} 
\rho\left(\{z, w\} + \b N \frac{(2\pi)^2}{\tA}  z w \right).
\end{eqnarray}
Therefore, the deformation 
(replacing the commutator by the $*$-commutator) amounts, in the continuum theory,
to replacing the Lie brackets $\{z,w\}$ as
\begin{equation}
\{z,w\} \longrightarrow \{z,w\}+\b N \frac{(2\pi)^2}{\tA} zw.
\label{RFContinuumStar1}
\end{equation}
Similarly, the Lie brackets
$\{z, \bw\}$ should be replaced as
\begin{equation}
\{z,\bw\} \longrightarrow \{z,\bw\}-\b N \frac{(2\pi)^2}{\tA} z \bw.
\label{RFContinuumStar2}
\end{equation}

The generalisation to the six-parameter deformation
is straightforward. We assume all deformation parameters $\b^{(I J)}$ to be of
order $1/N$. Then the deformation amounts to
\begin{eqnarray}
\{z^I,z^J\} &\longrightarrow& \{z^I,z^J\}
+
\b^{(I J)} N \frac{(2\pi)^2}{\tA} z^I z^J, 
\label{RFContinuumStarGeneral1}
\\
\{z^I,\bar{z^J}\} &\longrightarrow& \{z^I,\bar{z^J}\}
-
\b^{(I J)} N \frac{(2\pi)^2}{\tA} z^I \bar{z^J},
\label{RFContinuumStarGeneral2}
\end{eqnarray}
in the continuum theory.
Similar replacements are necessary 
for the Lie brackets between scalar fields and fermionic 
fields, which can be derived using projectors acting on fermionic
fields as in (\ref{RFStarcommutatorExampleScalarFermion}),
\begin{eqnarray}
\{z^I,\psi\} 
&\longrightarrow &
\{z^I,\psi\}+
\sum_J 
\b^{(IJ)} N
\frac{(2\pi)^2}{\tA}
z^I
\left(
\frac{1}{2} \c^{\bJ J} \psi
\right),
\label{RFContinuumStarGeneralFermion1}
\\
\{\bar{z^I},\psi\} 
&\longrightarrow &
\{\bar{z^I},\psi\} 
-
\sum_J 
\b^{(IJ)} N
\frac{(2\pi)^2}{\tA}
\bz^I
\left(
\frac{1}{2} \c^{\bJ J} \psi
\right).
\label{RFContinuumStarGeneralFermion2}
\end{eqnarray}
These will be used in \ref{RSSFermionic}.

\subsection{Background}
\label{RSSBackground}

Now that we know the continuum counterpart of the $*$-commutator,
it is easy to obtain the continuum theory which gives the deformed 
matrix model upon matrix regularisation;
one should just apply
the substitution (\ref{RFContinuumStar1}), (\ref{RFContinuumStar2})
and their complex conjugates to the Hamiltonian of the original 
continuum theory (\ref{RFHamiltonianFlatBosonic}).
Relevant terms in (\ref{RFHamiltonianFlatBosonic}) are
\begin{equation}
\int
\frac{\tA}{(-P_-)}
\left (
\{z,w\}\{\bz,\bw\}
+
\{z,\bw\}\{\bz,w\}
\right )
d^2\s,
\end{equation}
and we get the continuum version of the deformed matrix model,
\begin{eqnarray}
H&=&
\mbox{(orig.)}+
\int\Bigg( 
2 \frac{\tA}{(-P_{-})} \left(\b N \frac{(2\pi)^2}{\tA} \right)^2
z w \bar{z} \bar{w} 
+
\frac{\tA}{(-P_{-})} \left(\b N \frac{(2\pi)^2}{\tA} \right) 
\non\\
&\times&
\left( 
z w \{ \bz , \bw\} 
+
\bz \bw \{ z , w\} 
-
z \bw \{ \bz , w\} 
-
\bz w \{ z , \bw\} 
\right)
\Bigg) d^2 \s, 
\label{RFHamiltonianContinuumFromBeta}
\end{eqnarray}
where (orig.) stands for the original Hamiltonian (\ref{RFHamiltonianFlatBosonic}).
The constraint (\ref{RFConstraintAPD}) is left as it is.

Below, we shall show that this continuum theory described by
the Hamiltonian (\ref{RFHamiltonianContinuumFromBeta})
is identical to membrane theory on
a certain background.
The general action for the bosonic sector of membrane theory
coupled to the metric $G_{\m\n}(x)$  and the three-form gauge field 
$A_{\m\n\rho}(x)$ is given by,
\begin{eqnarray}
&&S=S_1+S_2,
\label{RFActionMembraneCurvedBosonicSum}
\\
&&S_1 =-T \int \sqrt{-\det (h_{ij})} \  d^2\s d\t,
\label{RFActionMembraneCurvedBosonicArea}
\\ 
&&h_{i j}= G_{\m \n} \fracpp{x^\m}{\s^i}\fracpp{x^\n}{\s^j}, \qquad
(i,j=0,1,2), 
\\
&&S_2=
T \int A_{\m\n\r}
\fracpp{x^\m }{\s^0} 
\fracpp{x^\n }{\s^1} 
\fracpp{x^\r }{\s^2}
d^3\s. \label{RFActionMembraneCurvedBosonicGauge}
\end{eqnarray}
We have put the membrane tension $T(=1)$ in 
(\ref{RFActionMembraneCurvedBosonicGauge}) 
to make $A_{\m\n\r}$ dimensionless.

Introduction of these backgrounds
affects the phase space constraint 
(\ref{RFConstraintTau}), (\ref{RFConstraintSigma}):
the flat metric 
is replaced by the curved metric,
and 
the three-form gauge field shifts the momenta
$\mP_\m$ to $\mP_\m - (1/2) A_{\m \n \r} \{x^\n, x^\r\}$.
Hence we have,
\begin{eqnarray}
&&G^{\m \n} 
(\mP_\mu - \frac{1}{2}A_{\mu \rho \s} \{x^\rho , x^\s\})
(\mP_\nu - \frac{1}{2}A_{\nu \lambda \t} \{x^\lambda , x^\t\})
\non\\
&&\quad + \frac{1}{2} 
G_{\m \rho}
G_{\nu \s}
\{x^\mu, x^\nu\}
\{x^\rho, x^\s\}
= 0, 
\label{RFConstraintTauCurved}
\\
&&
(\mP_\mu - \frac{1}{2}A_{\mu \nu \rho} \{x^\nu , x^\rho\})
\fracpp{x^\mu}{\s^r} =0, \qquad (r=1, 2). \label{RFConstraintSigmaCurved}
\end{eqnarray}
\Chk{**}

As we shall soon see,
it is sufficient to introduce 
only $G^{--}$ (or $G_{++}$) and  
$A_{+ \alpha \beta}$ components of
the background fields.
The curved background
of this type (often called the pp-wave or the plane-wave background)
is particularly well suited to
the lightcone gauge formalism;
the lightcone membrane theory on the background
can be derived simply by following the same steps 
as in the subsection \ref{RSSReviewFlatspaceMR},
starting from (\ref{RFConstraintTauCurved}) and
(\ref{RFConstraintSigmaCurved}).
The resulting Hamiltonian is,
\begin{eqnarray}
H&=&-P_{+}= \mbox{(orig.)}
\non\\
&+&
\int \left(
\frac{\tA}{2 (-P_-)} \left(\frac{-P_-}{\tA}\right)^2 G^{--}
-\frac{1}{2}A_{+\a\b}\{x^\a,x^\b\}
\right) d^2 \s.
\label{RFHamiltonianContinuumFromBKG}
\end{eqnarray}
\XXX{Interpretation?}
The constraint (\ref{RFConstraintAPD}) is not affected by
the introduction of the background.

The background fields  can now be
identified by comparing 
(\ref{RFHamiltonianContinuumFromBeta}) and 
(\ref{RFHamiltonianContinuumFromBKG}). 
The terms linear in $\b$ in (\ref{RFHamiltonianContinuumFromBeta})
match with 
the terms containing $A_{+\alpha\beta}$ in 
(\ref{RFHamiltonianContinuumFromBKG}),
and the term quadratic in $\b$ corresponds to 
the term involving $G^{--}$; 
we get,
\begin{eqnarray}
G^{--}&=& 4 (2 \pi )^4 \alpha^2 |z|^2|w|^2,
\\
A_{+\bz\bw}
&=&
- (2\pi)^2 \alpha z w,
\\
A_{+zw}
&=&
- (2\pi)^2 \alpha \bz \bw,
\\
A_{+\bz w}
&=&
 (2\pi)^2 \alpha z \bw,
\\
A_{+z \bw}
&=&
 (2\pi)^2 \alpha \bz w.
\end{eqnarray}
\Chk{***}
Here we have defined 
\begin{equation}
\a=\frac{T}{-P_-}\b N.
\end{equation}
The background fields depend on the deformation parameter $\b$
only through this combination.~\footnote{
We have restored the membrane tension $T$ to see
that $\a$ has the dimension of $(\mbox{length})^{-2}$.}
In other words, 
this rescaled parameter $\a$, rather than $\b$,
is the appropriate parameter 
to measure the deformation of the background from flat spacetime.
The gauge invariant four-form flux defined by
\begin{equation}
F_{\m\n\r\s}
=\partial_\m A_{\n\r\s} \pm\mbox{(cyclic permutations)},
\end{equation}
are given by
\begin{eqnarray}
F_{+ z w \bw}&=& 2(2\pi)^2 \alpha \bz,
\nonumber\\
F_{+ w z \bz}&=& -2(2\pi)^2 \alpha \bw,
\nonumber\\
F_{+\bz\bw w}&=& 2(2\pi)^2 \alpha z,
\nonumber\\
F_{+\bw\bz z}&=& -2(2\pi)^2 \alpha w.
\end{eqnarray}
\Rder{
\begin{eqnarray}
F_{+ z w \bw}&=& -2(2\pi)^2 \alpha \bz,
\nonumber\\
F_{+ w z \bz}&=& 2(2\pi)^2 \alpha \bw,
\nonumber\\
F_{+\bz\bw w}&=& -2(2\pi)^2 \alpha z,
\nonumber\\
F_{+\bw\bz z}&=& 2(2\pi)^2 \alpha w.
\end{eqnarray}
}
Thus, the four-form flux is not constant and
depends linearly on transverse coordinates.

\XXX{This is the bosonic sector but 
there is a kind of remnant from the full supersymmetric membrane theory.
Full supersymmetric theory cannot have the local fermionic symmetry
if the BKG does not satisfy EOM.
}

It is crucial to see whether these
background
fields satisfy equations of motion of eleven-dimensional supergravity.
As is well known, and 
as we shall explain in \ref{RSSFermionic} (see in particular
discussion around 
(\ref{RFSuperspaceConstraintsT})-(\ref{RFSuperspaceConstraintsZero})), 
the $\kappa$-symmetry of the supermembrane action
is related to the following
equations of motion for the background
\begin{eqnarray}
-{R_{\mu \nu \rho}}^{\mu}
&=&
-\frac{1}{12} 
F_{\nu \mu_1 \mu_2 \mu_3}
{F^{\mu_3 \mu_2 \mu_1}}_\rho
+
\frac{1}{144} 
F_{\mu_1 \mu_2 \mu_3 \mu_4}
F^{\mu_4 \mu_3 \mu_2 \mu_1}
G_{\nu \rho},
\label{RFEOMGGeneral}
\\
D_\m F^{\m\n\r\s}&=&0,
\label{RFEOMFGeneral}
\end{eqnarray}
\Rder{
\begin{eqnarray}
{R^\mu}_{\nu \rho \mu}
&=&
-\frac{1}{12} 
F_{\nu \mu_1 \mu_2 \mu_3}
{F^{\mu_3 \mu_2 \mu_1}}_\rho
+
\frac{1}{144} 
F_{\mu_1 \mu_2 \mu_3 \mu_4}
F^{\mu_4 \mu_3 \mu_2 \mu_1}
G_{\nu \rho}
\label{RFEOMMetricGeneralRder}
\\
D_\m F^{\m\n\r\s}&=&0
\end{eqnarray}
}
\Chk{**}
where we have omitted 
the Chern-Simon coupling 
$\e^{\n\r\s\m_1\cdots\m_5\t_1\cdots\t_5}
F_{\m_1\cdots\m_5}F_{\t_1\cdots\t_5}$
in (\ref{RFEOMFGeneral}),
as
it vanishes trivially for our background.
For our convention of the curvature tensor, see bosonic components of
(\ref{RFDefSuperCurvature}).
\XXX{Improve?}
These equations of motion
reduce for the pp-wave background to
\begin{eqnarray}
\frac{1}{2}\partial^\a \partial_\a (G^{--}) &=& \frac{1}{12} 
F_{+\a\b\c} {F_+}^{\a\b\c},
\label{RFEOMPPG}
\\
\partial^\a F_{+ \a\b\c} &=& 0.  
\label{RFEOMPPF}
\end{eqnarray}
\Chk{***}
\Rder{
\begin{equation}
R_{++}=-\frac{1}{2}\partial^\a \partial_\a (G^{--}) 
\end{equation}
}
Our background solves these equations of motion.
We wish to stress the strictness of this requirement.
Not only the forms of various components 
of the background fields,
but also the overall coefficient in (\ref{RFEOMPPG})
should be correct.
This high degree of consistency is achieved without 
any artificial tuning of parameters.
In particular,
the numerical factor $\frac{1}{12}$ in the equation of motion
for the metric (\ref{RFEOMGGeneral}) cannot
be absorbed into rescaling of $A_{\m\n\r}$ and $G_{\m\n}$,
since the normalisation convention 
of them is already fixed 
by choosing the membrane action to be 
(\ref{RFActionMembraneCurvedBosonicSum})-(\ref{RFActionMembraneCurvedBosonicGauge}).~\footnote{
The special rescaling of the background fields
$G_{\mu\nu}^\prime=
\lambda G_{\m\n}$ and 
$A_{\m\n\r}^\prime
=\lambda^{3/2}A_{\m\n\r}$ only changes the action by
an overall factor, which can be absorbed 
into a redefinition of the tension $T$.
Hence this rescaling should not and does not change the physics of the background.
In particular, it does not affect the equation of motion 
(\ref{RFEOMGGeneral}). We have fixed this rescaling 
by choosing the components of the metric other than $G_{++}$
to be equal to those of the flat spacetime metric.
}

For the general six-parameter deformation,
we should apply the substitution
(\ref{RFContinuumStarGeneral1})(\ref{RFContinuumStarGeneral2}) to
the Hamiltonian
(\ref{RFHamiltonianFlatBosonic}),
and again compare it to the expression (\ref{RFHamiltonianContinuumFromBKG}).
By using the parameter $\alpha^{(I J)}$ for the continuum theory defined by,
\begin{equation}
\a^{(IJ)} = \frac{T}{-P_-} \b^{(IJ)}  N,
\label{RFAlphaDefGen}
\end{equation}
the background thus identified is given by,
\begin{eqnarray}
G^{--}&=&2(2 \pi)^4   \sum_{I}\sum_{J} \left(\a^{(IJ)}\right)^2 |z^I|^2|z^J|^2,
\label{RFBKGGeneralG}
\\
A_{+I\bJ}&=& (2\pi)^2 \alpha^{(IJ)} \bar{z^I} z^J,\\
A_{+IJ}&=&- (2\pi)^2 \alpha^{(IJ)} \bar{z^I} \bar{z^J},\\
A_{+\bI\bJ}&=&- (2\pi)^2 \alpha^{(IJ)} z^I z^J,
\end{eqnarray}
\Chk{**}
with the four-form flux
\begin{eqnarray}
F_{+IJ\bJ}&=&2(2\pi)^2\alpha^{(IJ)} \bar{z^I},
\label{RFBKGGeneralF1}
\\
F_{+\bI \bJ J}&=&2(2\pi)^2\alpha^{(IJ)} z^I.
\label{RFBKGGeneralF2}
\end{eqnarray}
\Rder{
\begin{eqnarray}
F_{+IJ\bJ}&=&-2(2\pi)^2\alpha^{(IJ)} \bar{z^I}\\
F_{+\bI \bJ J}&=&-2(2\pi)^2\alpha^{(IJ)} z^I.
\end{eqnarray}
}
These background fields again solve the equations of motion
(\ref{RFEOMPPG})(\ref{RFEOMPPF}). 
\XXX{Note AR=AL FR=-FL, when all indices are Bosonic}

\subsection{Fermionic Sector}
\label{RSSFermionic}
So far, we have focused on bosonic degrees of freedom,
and have identified the background 
(\ref{RFBKGGeneralG})-(\ref{RFBKGGeneralF2}).
We will now consider the fermionic sector of supermembrane theory
propagating on this background, and show that
it precisely reproduces the fermionic sector of the deformed matrix model; 
the prescription for the deformation,
introduced in section \ref{RSDeformation}, 
is consistent with the eleven-dimensional physics, 
even including the fermionic sector.

Let us first recall some of the basic properties of
supermembrane theory on curved 
backgrounds~\cite{RBBergshoeffSezginTownsendShort}\cite{RBBergshoeffSezginTownsendLong}.
For brevity, we will frequently refer the reader to appendix \ref{RSASuper}
for explicit formulae. 
There is a 32-component fermionic field on the membrane worldvolume,
$\th^a(\s^0,\s^1,\s^2)$, $a=1,\cdots, 32$. 
We write 
$x^\mu$ and $\th^a=x^a$ collectively as $x^A$, $A=(\mu, a)$,
which can be considered as coordinates on a superspace.
The full action describing the supermembranes on general curved 
backgrounds is given by,
\begin{eqnarray}
&&S=S_1+S_2,
\label{RFActionMembraneCurvedSuperTotal}
\\
&&S_1=
- \int \sqrt{-\det (h_{ij})} \  d^2\s d\t,
\label{RFActionMembraneCurvedSuperArea}
\\ 
&&h_{i j} = \eta_{\hmu \hnu} {\pi_i}^\hmu {\pi_j}^\hnu,
\qquad
(i,j=0,1,2), 
\\
&&
{\pi_i}^\hmu = \partial_i x^A {E_A}^\hmu,
\\
&&S_2=
\int \partial_2 x^C \partial_1 x^B \partial_0 x^A
A_{ABC} d^3\s.
\label{RFActionMembraneCurvedSuperGauge}
\end{eqnarray}

\Rder{
\begin{eqnarray}
\\&& {\pi^\hmu}_i = {E^\hmu}_A {x^A}_{,i}
\\&& S_2 = \int A^{(R)}_{ABC} x^C_{,2} x^B_{,1} x^A_{,0} d^3\s.
\end{eqnarray}
}
Only in this section, we distinguish tangent-space indices
$\hA=(\hmu, \ha), \hB=(\hnu, \hb), \cdots$ from curved-space indices $A,B,\cdots$.
The background superfields in this action are (a part of) 
the supervielbein  ${E_A}^\hmu(x^\nu, \th^a)$ and the three-form
gauge potential $A_{ABC}(x^\mu, \th^a)$.
If we take $\th=0$, 
${E_\nu}^\hmu$ 
and 
$A_{\m\n\r}$ reduce to the bosonic component fields 
of supergravity,
the elfbein and the three-form gauge field.
For our background, the gravitino field is not present.

A key feature of supermembrane theory is 
the 
$\kappa$-symmetry 
(\ref{RFKappaV})(\ref{RFKappaS}), which is 
the local fermionic symmetry 
with 16 anti-commuting parameters.
The action is $\kappa$-symmetric if 
the following constraints on the superspace torsion and 
the field strength tensor are satisfied,
\footnote{One might wonder why these constraints
include (through the definition of the torsion) superfields which are not contained
in the action, i. e. the vielbein with spinor tangent index, ${E_A}^\ha$,
and the connection,
${\O_{A\hB}}^\hC$.
An answer to this question is that these extra superfields act as kinds of
integration constants: 
the action described by ${E_A}^\hmu$ and $A_{ABC}$
is $\kappa$-symmetric  
when ${E_A}^\ha$ and ${\O_{A\hB}}^\hC$ exist
such that, together with given 
${E_A}^\hmu$ and $A_{ABC}$,
(\ref{RFSuperspaceConstraintsT})-(\ref{RFSuperspaceConstraintsZero}) are satisfied.
Similar issues for superstring theory are discussed in
\cite{RBTaylorShapiro}.
We also remark that in 
\cite{RBBergshoeffSezginTownsendShort}\cite{RBBergshoeffSezginTownsendLong}
apparently weaker constraints are given, 
which are equivalent to 
(\ref{RFSuperspaceConstraintsT})-(\ref{RFSuperspaceConstraintsZero})
by a field 
redefinition~\cite{RBDuffHoweInamiStelle}\cite{RBBergshoeffSezginTownsendLong}.  }
\begin{eqnarray}
&&
{T_{\ha\hb}}^\hmu
=
-2i 
{\C^\hmu}_{\ha\hb},
\label{RFSuperspaceConstraintsT}
\\
&&
F_{\hmu\hnu\ha\hb} 
=\kpm 2i \C_{\hmu\hnu\ha\hb},
\label{RFSuperspaceConstraintsF}
\\
&&
0
={T_{\hmu\hnu}}^\hrho
={T_{\ha\hmu}}^\hnu
=F_{\hmu\hnu\hrho\hb}
=F_{\hmu\ha\hb\hc}
=F_{\ha\hb\hc\hd}.
\label{RFSuperspaceConstraintsZero}
\end{eqnarray}
\Rder{
\begin{eqnarray}
&& {T^\hmu}_{\ha\hb} =s_1^R {\C^\hmu}_{\ha\hb} = 2i {\C^\hmu}_{\ha\hb} \\
&& F^{(R)}_{\hmu\hnu\ha\hb} 
= s^{(R)}_2 \C_{\hmu\hnu\ha\hb}  
= \kmp 2 i \C_{\hmu\hnu\ha\hb}.
\\
\end{eqnarray}
}
Our conventions for the $32\times 32$
gamma matrices $\C^\hmu$,
the superspace torsion $T$,
and the four-form field strength $F$
are summarised in
appendix \ref{RSASuper},
(\ref{RFGammaAC32})-(\ref{RFGammaMinus}),
(\ref{RFGammaLL}), (\ref{RFDefTorsion})-(\ref{RFDefSuperFHat}).

These constraints are equivalent to the fundamental equations
in the superspace formulation of 
eleven-dimensional supergravity~\cite{RBBrinkHowe}\cite{RBCremmerFerrara}.
In this way, the information of the equation of motion 
of eleven-dimensional supergravity is incorporated in supermembrane theory. 
The component formulation of supergravity,
which was used in subsection \ref{RSSBackground},
is related to the superspace formulation in the following way.
The equations of motion for the component supergravity fields
are derived in  
\cite{RBBrinkHowe}\cite{RBCremmerFerrara}
from the superspace constraints 
(\ref{RFSuperspaceConstraintsT})-(\ref{RFSuperspaceConstraintsZero})
by successive applications of Bianchi identities 
(\ref{RFBianchiDT})-(\ref{RFBianchiDF})
fixing, in particular, the numerical coefficients in (\ref{RFEOMGGeneral}).
Thus, the superspace formulation
implies the component formulation.
It is believed that the converse is also true:
given component fields satisfying the component equations of motion,
it is widely assumed that superfields exist which satisfy
the conditions,
(a) their lowest non-trivial components coincide with the given component
fields, (b) they satisfy all of the superspace constraints
(\ref{RFSuperspaceConstraintsT})-(\ref{RFSuperspaceConstraintsZero}),
order by order in the $\theta$-expansion.
Although this property is not proven,
we shall also assume this here, as it is highly unlikely 
that the two formulations are not equivalent,
because of the strong restriction 
from the local supersymmetry.

The full construction of the superfields from given
component fields 
is also technically hard in general,
and is so far achieved only for special cases 
with a high degree of symmetry (see e.~g.~\cite{RBSuperFieldsCoset}). 
Instead of constructing the full superfields
for our background,
we will directly 
obtain the Hamiltonian in the lightcone gauge,
just by assuming the existence of the full superfields.
This is possible
because,
for the pp-wave background, 
the gauge fixing condition for the $\k$-symmetry,
\begin{equation}
{\left(\C^{\hat{+}}\right)^\ha}_{\hb} \theta^\hb = 0,
\qquad
\th^\ha=\delta^\ha_b \th^b,
\label{RFGaugeConditionTheta}
\end{equation}
drastically reduces
the number of relevant terms appearing 
in the Hamilton formalism, as we shall explain below.
This type of argument is used for example in 
\cite{RBMetsaevTseytlin}\cite{RBRussoTseytlin} for type IIB string theory.
Our treatment more closely follows that in 
\cite{RBMizoguchiMogamiSatoh}.
See \cite{RBIizuka} for the application 
to supermembrane theory
on the pp-wave background with a constant
flux.

We start by observing that,
in the $\th$-expansion of a superfield,
any pair of two $\th$'s
can be written in terms of 
fundamental bi-spinors $ {\C^{\hmu_1 \cdots \hmu_n}}_{\ha\hb}\th^\ha \th^\hb$ 
with $n=1,2,5$.
Under the condition (\ref{RFGaugeConditionTheta}),
only non-zero bi-spinors are those with
single upper $\hat{-}$ index with arbitrary number 
of transverse SO(9) vector indices,
${\C^{\hat{-}\halpha_1\cdots\halpha_{n-1}}}_{\ha\hb}\th^\ha \th^\hb$.~\footnote{
In general, for any two spinors $\xi,\eta$ satisfying
$\Gamma^{\hat{+}} \xi =0 , \Gamma^{\hat{+}} \eta=0$,
the expression ${\Gamma^{\hmu_1\cdots\hmu_n}}_{\ha\hb} \xi^\ha \eta^\hb$ vanishes
except for 
${\Gamma^{\hat{-}\halpha_1\cdots\halpha_{n-1}}}_{\ha\hb} \xi^\ha \eta^\hb$.
}

Another necessary ingredient is a gauge fixing for the
background superfields~\cite{RBNorcor},
similar to the normal coordinates in Riemannian geometry.
The gauge transformations
for the backgrounds, namely, the general coordinate
transformation on the superspace, the local Lorentz transformation,
and the gauge transformation for the three-form field,
are (partially) fixed by imposing the conditions 
(\ref{RFNorcorGC})-(\ref{RFNorcor3form}).
In this gauge, 
it is possible to formulate an algorithm
which iteratively calculates higher order terms
in the $\th$-expansion, based on
a part (but not all) of the constraints and Bianchi identities.
It is known that,
as a result,
the coefficients in the $\th$-expansion
in this gauge are expressed
in terms of ${R_{\hmu\hnu\hrho}}^\hsigma$, 
$F_{\hmu\hnu\hrho\hsigma}$, 
${\O_{\hmu\hnu}}^\hrho$, ${E_\mu}^\hnu$,
and their covariant derivatives in the bosonic directions,
evaluated at $\th=0$.
The vector indices of the fundamental bi-spinors
should be contracted with these structures,
except for the indices of the original superfield.

However, on the pp-wave background 
(which does not depend on $x^+$),
these expressions are trivial except for
${E_+}^{\hat{-}}|_{\th=0}=\frac{1}{2}G^{--}$,
$\O_{+ \a +}|_{\th=0}= \frac{1}{2} \partial_\a 
G^{--},
R_{+\a+\b}|_{\th=0}=
-\frac{1}{2} \partial_\a  \partial_\b G^{--}$,  
\Chk{-}
and their covariant derivatives
in the transverse directions.
Thus, there are no
lower $\hat{-}$ indices to match
the upper $\hat{-}$ indices coming from 
the bi-spinors.
Hence, only a few terms in the $\th$-expansion
of various superfields
can survive in the lightcone gauge formulation 
of supermembrane theory on the pp-wave background.
In fact, 
one can show, with the help of some dimensional analysis, that
there are only three relevant terms,
except for the purely bosonic ones already treated in 
\ref{RSSBackground}.
These relevant terms are 
the $(\th)^1$-part
(linear in $\th$'s) of 
$A_{\mu \nu a}$ and 
${E_a}^\hmu$,
and 
the $(\th)^2$-part of 
${E_+}^{\hat{-}}$.
The first two terms exist for 
flat spacetime and are unchanged 
for our background.
They are respectively responsible for (the commutator part of) 
the Yukawa couplings 
and the Dirac brackets for the fermionic variables in the matrix model.
The third term vanishes for flat spacetime,
and is the only new contribution from fermionic fields,
appearing in the curved background.
We shall see below that this term also contributes to the Yukawa couplings
of the matrix model, and deform the commutators into $*$-commutators.
Other terms either vanish by themselves or 
do not appear in the lightcone formalism,
due to the relations, $\fracpp{x^+}{\s^r}=0\ (r=1,2)$, and
$\c^+ \fracpp{\th}{\s^i}=0\ (i=0,1,2)$.

Explicit expressions for these three terms can be calculated
using the gauge fixing condition~\cite{RBNorcor}
\begin{eqnarray}
\th^b \left.\partial_b A_{\mu \nu a}\right|_{\th=0}
&=&
\th^b
\frac{1}{2} F_{b a \m \n}|_{\th=0}
=
\kpm i
\th^\hb 
\C_{\mu \nu \hb \hc} 
\delta^\hc_a,
\label{RFNorcorAVVS1}
\\
\th^b\partial_b{E_a}^\hmu|_{\th=0}
&=&
-i \th^\hb {\c^\hmu}_{\hb\hc} \d^\hc_a,
\label{RFNorcorESV1}
\\
\frac{1}{2!} \th^b \th^a \left. \partial_a \partial_b {E_\nu}^{\hmu} 
\right|_{\th=0}
&=&
\kpm \frac{i}{36}
\th^\hb  \th^\hc
\Bigg({\C^\hmu}
\Bigg
(F_{\nu\hsigma_1\hsigma_2\hsigma_3} \C^{\hsigma_3 \hsigma_2 \hsigma_1}
\non\\
&&
+\frac{1}{8} F_{\hsigma_1\hsigma_2\hsigma_3\hsigma_4}
{\C^{\hsigma_4 \hsigma_3 \hsigma_2 \hsigma_1}}_\nu\Bigg)
\Bigg|_{\th=0}\Bigg)_{\hb \hc}.
\label{RFNorcorEVV2}
\end{eqnarray}
These expressions are also derived using another method,
the method of the gauge completion,
in \cite{RBdeWitPeetersPlefka}\cite{RBShibusa}. 

\Rder{
\begin{equation}
{E^\hmu}_{a,b}\th^b
=\frac{1}{2}{T^\hmu}_{ab}\th^b
=i {\c^\hmu}_{\hc\hb} \th^b {\d^\hc}_a
\end{equation}
}
\Rder{
\begin{eqnarray}
&&\left.{E^\hmu}_{\nu,a,b}\right|_{\th=0} \frac{1}{2!} \th^b \th^a
=
-\frac{1}{2} \left. \left({T^\hmu}_{\hb \ha} {T^\ha}_{\hc \hrho}
{E^\hrho}_\nu\right)\right|_{\th=0}  \th^\hb \th^\hc
\\
&&=\frac{1}{72}\frac{s_1}{s_2}s_1
\th^\hb 
{
\left({\C^\hmu}
\left
(F_{\nu\hsigma_1\hsigma_2\hsigma_3} \C^{\hsigma_3 \hsigma_2 \hsigma_1}
+\frac{1}{8} F_{\hsigma_1\hsigma_2\hsigma_3\hsigma_4}
{\C^{\hsigma_4 \hsigma_3 \hsigma_2 \hsigma_1}}_\nu\right)\right)}_{\hb\hc}
 \th^\hc,
\end{eqnarray}
}
\Rder{
\begin{equation}
{A_{\mu\nu a, b}}\bigg|_{\th=0} \th^b
=\left. \frac{1}{2}F_{\mu\nu a b}\right|_{\th=0} \th^b
=\frac{1}{2} s_2 \C_{\mu\nu \ha \hb} {\delta^\ha}_a \th^\hb
\end{equation}
}

The lightcone gauge formulation of supermembrane theory 
on our background can now be derived, 
in a way similar to the bosonic theory.
The starting point is phase space constraints,
(\ref{RFConstraintTauSuper})(\ref{RFConstraintSigmaSuper}),
which are supersymmetric generalisations of
(\ref{RFConstraintTauCurved})(\ref{RFConstraintSigmaCurved}),
and a new constraint (\ref{RFConstraintFermionicMomenta}), 
which solves canonical momenta of $\th$ completely in terms of
other variables.
We shall skip most of the intermediate steps and just
present the result of this analysis in the following.

Firstly, the Dirac brackets can be calculated in the standard way, 
using the condition (\ref{RFGaugeConditionTheta})
and the constraint (\ref{RFConstraintFermionicMomenta}), with the help of 
(\ref{RFNorcorESV1}),
\begin{equation}
\sqrt{2} \frac{(-P_-)}{\tA}
\{
\psi^a(\s')
,
\psi^b(\s'')
\}_{\mbox{D.B.}}
=
-\frac{i}{2} \d^{ab} \d^2(\s'-\s'').
\label{RFDBFermionMembrane}
\end{equation}
For the supersymmetric case, 
$\partial_r x^-$ also has the contribution
from the fermionic coordinates,
in addition to the right-hand side of (\ref{RFBosonicSolvedXm}).
Consequently, the constraint
corresponding to the area preserving diffeomorphism becomes, 
using (\ref{RFNorcorESV1}),
\begin{equation}
0=
\{
\mP^\a, x^\a
\}
+
\frac{-P_-}{\tA}
i \sqrt{2}
\{
\psi^T, \psi
\}.
\label{RFConstraintAPDSuper}
\end{equation}

We now focus on the Hamiltonian. The right-hand side of
(\ref{RFNorcorEVV2}) reduces to
\begin{equation}
\kmp \frac{\sqrt{2}i}{24} 
F_{+\halpha_1\halpha_2\halpha_3}
\psi^T
\c^{\halpha_3\halpha_2\halpha_1}
\psi,
\label{RFNorcorEVV2LC}
\end{equation}
\Chk{-}
where $\psi$ is the 16-component spinor,
defined in (\ref{RFDef16Spinor}),
which is a part of the 32-component spinor $\th$ 
surviving the lightcone gauge condition (\ref{RFGaugeConditionTheta}).
This term contributes to the lightcone Hamiltonian
in a similar manner as the first term 
in (\ref{RFHamiltonianContinuumFromBKG}),
through the relation $G^{--}=2 {E_+}^{\hat{-}}$.
Another contribution to the Hamiltonian comes from
(\ref{RFNorcorAVVS1}), in a similar way to 
the second term in (\ref{RFHamiltonianContinuumFromBKG}),
\begin{equation}
\int A_{+\nu a}\{x^\nu,\th^a\} d^2\s
=\int \kpm \sqrt{2}i
\psi^T \c^\a \{x^\a,\psi\} d^2\s.
\end{equation}
We note that $A_{+\nu a}$ is anti-commuting.

\Rder{
\begin{equation}
(\th)^2\mbox{-part of }{E^{\hat{-}}}_+=
-\frac{1}{48} \sqrt{2} \frac{s_1}{s_2} s_1
F_{+\halpha_1\halpha_2\halpha_3}
\psi^T
\c^{\halpha_3\halpha_2\halpha_1}
\psi
\end{equation}
\begin{equation}
=
\kpm \frac{\sqrt{2} i}{24}
F_{+\halpha_1\halpha_2\halpha_3}
\psi^T
\c^{\halpha_3\halpha_2\halpha_1}
\psi
\end{equation}
\begin{eqnarray}
\mbox{contribution to }-\mP_+
=
\int - A_{+\nu a}\{x^\nu,\th^a\} d^2\s
=\int -\frac{\sqrt{2}}{2}s_2^R
\psi^T \c^\a \{x^\a,\psi\} d^2\s
\end{eqnarray}
\begin{equation}
=\int \kpm \sqrt{2}i
\psi^T \c^\a \{x^\a,\psi\} d^2\s
\end{equation}
}

We thus obtain the full Hamiltonian
for supermembranes propagating on our background,
\begin{eqnarray}
&&H=\mbox{(bosonic.)}+
\int
\Bigg( 
\kpm \sqrt{2}i
\psi^T \c^\a \{x^\a,\psi\}
\non\\
&\kmp&
\frac{\sqrt{2} i }{2}
(2\pi)^2
\frac{-P_-}{\tA}
\sum_{I,J}
\a^{(IJ)}
\left(
\bz^I \psi^T \c^{\bJ J I} \psi
+
z^I \psi^T \c^{ J \bJ \bI} \psi
\right) 
\Bigg)
d^2\s,
\label{RFHamiltonianSuperFromBKG}
\end{eqnarray}
where $\mbox{(bosonic.)}$ stands for 
the purely bosonic part of the Hamiltonian 
for our background.

We should compare this 
with the continuum theory
corresponding
to the deformed matrix model (\ref{RFHamiltonianDeformedMM}), 
in the regime $\b N \sim1$.
As we have seen for the bosonic sector,
this continuum theory can be obtained 
by deforming the Lie brackets
of the original continuum theory for flat spacetime. 
For the fermionic sector, 
the Hamiltonian for flat spacetime is given by
the first term in the integrand of (\ref{RFHamiltonianSuperFromBKG}),
and the relevant substitution is 
(\ref{RFContinuumStarGeneralFermion1})(\ref{RFContinuumStarGeneralFermion2}).
We get,
\begin{eqnarray}
&&H=\mbox{(bosonic.)}+
\int \Bigg(\kpm \sqrt{2}i
\psi^T \c^\a \{x^\a,\psi\} 
\non\\
&+&
\kpm
\sqrt{2} i
\psi^T
\sum_{I,J}
\Bigg(
\c^{\bI}
\b^{(IJ)}N
\frac{ (2\pi)^2}{\tA}
z^I 
\left(
\frac{1}{2}
\c^{\bJ J} \psi 
\right)
\non\\
&& \qquad \qquad \qquad \qquad
-
\c^{I}
\b^{(IJ)}N
\frac{ (2\pi)^2}{\tA}
\bar{z^I} 
\left(
\frac{1}{2}
\c^{\bJ J} \psi 
\right)
\Bigg)
\Bigg)
d^2\s.
\end{eqnarray}
This expression precisely matches with (\ref{RFHamiltonianSuperFromBKG}),
under the definition (\ref{RFAlphaDefGen}). 
Thus, we have shown that
the deformed matrix model (in the scaling limit)
is equivalent to
matrix-regularised supermembrane theory 
on our background.

\section{Stable solution}
\label{RSSolution}
In this section, we consider some 
stable solutions in the deformed model.
They correspond to membranes with torus topology.
We also discuss some of their properties.
In particular, we shall show that two 
apparently distinct
configurations of membranes, labelled by different winding numbers,
are actually equivalent in the matrix model.
We also consider classical flat directions associated with the solutions,
and discuss quantum corrections to them,
using an analogy to the four-dimensional theory.

We shall focus on the single-parameter deformation 
for simplicity.~\footnote{
By considering the general 
deformation it should be possible to construct
higher dimensional analogues of the stable solutions
considered here.
}
We first observe that
every zero-energy configuration is a 
(marginally) stable configuration,
since the potential term is always non-negative.
Therefore, if one has a configuration 
in which every $*$-commutators vanish,
the configuration is stable.
This is similar to the situation in the original matrix model
where a configuration with commuting (or 
simultaneously diagonalisable) $X^\a$ is a stable solution. 
For simplicity, we set all coordinates other than $Z$ and $W$
to zero.
Then the vanishing of all $*$-commutators  amounts simply to
\begin{equation}
[Z, W]_*=0, \quad [Z,W^\dagger]_*=0.
\label{RFVanishingStarComm}
\end{equation}

If the deformation parameter $\beta$ takes one of the special values,
$\beta= \frac{n}{N}$, where $n$ is an integer,
this equation can be solved explicitly
\footnote{
An four-dimensional analogue of this class of solutions
is first discussed in \cite{RBBerensteinLeigh}.
See also \cite{RBLuninMaldacena}.}
by using 
the so-called clock, shift matrices defined by,
\begin{eqnarray}
h_1=\left(
\begin{array}{cccc}
1 &                      &         &0 \\
  & e^{i\frac{2 \pi}{N}} &         &  \\
  &                      & \ddots  &  \\
0 &                      &         &e^{i\frac{2 \pi}{N}(N-1)}
\end{array}
\right),
h_2=
%
\left(
\begin{array}{cccc}
0 &  1     &        & 0\\
  & \ddots & \ddots &  \\
  & 0      & \ddots & 1\\
1 &        &        & 0 
\end{array}
\right),
\end{eqnarray}
which satisfy
\begin{equation}
h_1 h_2 = e^{- i \frac{2 \pi}{N}} h_2 h_1.
\end{equation}
\Chk{*}
In the simplest case, $\b=\frac{1}{N}$,
\begin{equation}
Z= a h_1, \quad W= b h_2,
\label{RFSolutionMMSimplest}
\end{equation}
\Chk{*}
is a solution to (\ref{RFVanishingStarComm})
where $a, b$ are arbitrary parameters.

As is well known, the matrices $h_1, h_2$ 
play a basic role in matrix regularisation of membranes
with torus topology~\cite{RBMRTorus}. A function on a 
torus can be represented
by a function 
defined on $[0,2\pi]\times[0,2\pi]$,
periodic in both $\s^1$, $\s^2$ directions. 
In this convention, the matrices $h_1$, $h_2$ correspond to the functions 
\begin{equation}
e^{i\s^1}, \quad e^{i\s^2},
\end{equation}
respectively. 
\Chk{-} 
All periodic functions can be generated from them,
and the function $e^{i (m_1 \s^1+ m_2 \s^2)}$
corresponds to,
\begin{equation}
e^{i \frac{\p}{N} m_1 m_2} h_1^{m_1} h_2^{m_2},
\end{equation}
where the extra phase factor ensures the correct behaviour 
under the complex conjugation.
Thus, the stable solution (\ref{RFSolutionMMSimplest})
corresponds to a configuration in  membrane theory
\begin{equation}
z=a' e^{i\s^1}, \quad w=b' e^{i \s^2},
\end{equation}
where  arbitrary constants $a', b'$ are given by
$a'=(2 \pi T)^{-\frac{1}{3}}a$,
$b'=(2 \pi T)^{-\frac{1}{3}}b$,
using (\ref{RFRescalingX}).
This configuration 
describes
a membrane with torus topology,
which are embedded into four-dimensional space
(parametrised by $x^1,x^2,x^3,x^4$)
with a simple $S^1\times S^1$ shape.
It is easy to check that this is a solution to continuum membrane theory 
(\ref{RFHamiltonianContinuumFromBKG})
on our background,
noting the relation
$\tA=(2\p)^2$.

In general, for $\b=\frac{n}{N}$ with any integer $n$,
the following matrices are solutions to the equation
(\ref{RFVanishingStarComm}),
\begin{equation}
Z = a e^{i \frac{\p}{N} l_1 l_2} h_1^{l_1} h_2^{l_2},\quad
W = b e^{i \frac{\p}{N} m_1 m_2} h_1^{m_1} h_2^{m_2},
\label{RFStableSolMatrixGeneral}
\end{equation}
when 
four integers
$l_1,l_2, m_1, m_2$ 
satisfy
\begin{equation}
l_1 m_2 - l_2 m_1 = n.
\end{equation}
The matrices (\ref{RFStableSolMatrixGeneral}) correspond, 
in the continuum theory, to
\begin{equation}
z= a' e^{i (l_1 \s^1+ l_2 \s^2)}, \quad
w= b' e^{i (m_1 \s^1+ m_2 \s^2)}, 
\label{RFStableSolMembraneGeneral}
\end{equation}
describing a membrane
which is in general wrapped on the same $S^1\times S^1$ several times.

For given $n$, some of the membrane solutions 
(\ref{RFStableSolMembraneGeneral}) describe the same
object in a different parametrisation, e. g.
$z= a' e^{i \s^1}, w= b' e^{i(\s^2 + \s^1)}$
and
$z= a' e^{i \s^1}, w= b' e^{i\s^2}$. 
\Chk{-}
The corresponding matrix solutions are equivalent 
by some unitary transformation, as it should be.
On the other hand, some of these membrane configurations
are physically distinct,
in the conventional membrane picture.
For example, for $n=2$, we consider the two configurations,
(a) $z=a' e^{i 2 \s^1}, w=b' e^{i \s^2}$ and 
(b) $z=a' e^{i \s^1}, w=b' e^{i 2 \s^2}$.
\Chk{-}
They have different winding numbers:
the case (a) corresponds to a membrane wrapped twice around
the circle in the $z$-plane (with radius $a'$) and once around that in the 
$w$-plane
(with radius $b'$), and (b) corresponds to a membrane
wrapped once around the circle in the $z$-plane and twice around that in 
the $w$-plane.
In conventional formulation of membrane theory, 
although they have the same energy, they are distinct objects.
In particular, they have different spectrums
for the fluctuations around them (except for the
special case $a'=b'$).
In (a), 
the allowed wave length of the fluctuations around
the configuration is 
$\frac{4 \pi a'}{l}$ in the $\s^1$ direction and 
$\frac{2 \pi b'}{m}$ in the $\s^2$ direction with integers 
$l,m$;
because of the double wrapping,
the fluctuation in the $s^1$ direction
allows excitation with doubled wave length 
$4\pi a'$.
For (b) the allowed wave length are 
$\frac{2 \pi a'}{l}$ and $\frac{4 \pi b'}{m}$.

However, one can show that the corresponding matrices
\begin{eqnarray}
\mbox{(a)}&:& 
Z= a (h_1)^2, \quad W= b h_2,
\\
\mbox{(b)}&:& 
Z= a h_1, \quad W= b (h_2)^2,
\end{eqnarray}
are related by a similarity transformation
$X^\a \rightarrow U X^\a U^{-1}$, with a unitary matrix $U$. 
Therefore, in the matrix model, these two configurations should
be considered physically equivalent.~\footnote{
Similar degeneracy is also noted in 
\cite{RBABHHS}.
}

In order to further understand this remarkable phenomena,
it is natural to focus on the fluctuation spectrums around these configurations 
in the matrix model, since, as is explained above,
in the conventional membrane picture, they  distinguish 
the two configurations.
We have computed the fluctuation
spectrums around these configurations,
and found that they are the same (as a matter of course) and
labelled by the ``wavelength''
$\frac{4 \pi a'}{l}$ and
$\frac{4 \pi b'}{m}$; in contrast to the membrane analysis,
the largest ``wavelength'' are 
doubled both in the $\s^1$ and $\s^2$ directions. 
This is technically a consequence of the appearance of a $\sin$-function 
instead of a linear function which
occurs in general for a discretised system.
The details will be presented elsewhere.

One possible interpretation to this remarkable degeneracy is the following.
It has long been  suspected that the membranes in M-theory are
non-Abelian objects, similar to D-branes. For recent interesting
developments, see e. g. \cite{RBLambertBagger}\cite{RBGustavsson}.
This non-Abelian nature might be making the
concept of winding numbers ill-defined.
For D-branes, one can argue that, for example,
at least the distinction is vague 
between 
(i) two coinciding D-branes, each of them wrapping a circle once
and (ii) single D-brane wrapping the circle twice. 
Starting from (i), we know that the 
coordinates are described by $2\times 2$ matrices X,
and the natural boundary condition for them, representing the wrapping,
is $X(0) = U X(2\pi) U^{-1}$ with unitary matrix $U$. 
If $U=1$ this gives the usual two singly wrapped D-branes.
However if one take $U=\s^1$, where $\s^1$ here represents the Pauli matrix,
this boundary condition describes single object wrapped around the circle twice.
The use of this type of boundary conditions plays an essential role in the matrix string
proposal~\cite{RBMatrixString1}\cite{RBMatrixString2}\cite{RBMatrixStringAmplitude}.
 
Another interesting aspect of these solutions concerns 
the parameters $a, b$ of them.
They are the radii of the two circles in the $z, w$-plane,
and can take arbitrary values. Thus, they parametrise the flat directions 
of the classical potential. The flat directions exist because,
in the lightcone gauge the force coming from 
the membrane tension is given by the usual double commutator term,
which is proportional to the cubic power of the coordinates; this can be balanced 
by the force from the quartic potential from the metric
(\ref{RFBKGGeneralG}), which also has the cubic dependence 
on the transverse coordinates. 
One can exploit these flat directions
to construct solutions which corresponds to two (or more) membranes
having different $a, b$, by arranging the solutions
corresponding to each membranes into 
block-diagonal matrices.
\XXX{Think! P-, N}

So far our discussion has been concrete.
Before concluding this section,
we would like to 
discuss, somewhat speculatively,
the quantum effect
to the classical flat directions, using the analogy
to four-dimensional $\b$-deformed $\mN=4$ SYM,
as this might give a very challenging application
of our deformed model.
In four-dimensional $\mN=4$ SYM,
the quantum corrections
do not break the classical scale invariance,
because of the cancellation
between bosonic and fermionic contributions,
which is presumably a consequence of 
the $\mN=4$ supersymmetry.
The $\b$-deformed SYM, while breaking the supersymmetry
(either down to $\mN=1$ or completely), preserves
the cancellation between fermionic and bosonic contributions.
This suggests that, although our deformed matrix model
in general is not supersymmetric, the bosonic and fermionic
contributions to the classical flat directions might cancel
each other.
\XXX{deWitLushcerNicolai?}
Furthermore, it is known that, for four-dimensional 
$\b$-deformed SYM, one has to introduce $1/N$-corrections
to various couplings, in addition to phase factors
from $*$-products,
in order to retain the boson-fermion cancellation.
One might expect that similar $1/N$ corrections are necessary
to make the classical flat directions flat even at the quantum level.
If this is  the case, this would give a new approach
to the long standing problem of the large $N$ limit of the matrix model;
by requiring the cancellation of the quantum corrections
to the flat directions, one might obtain information about
the behaviour of the couplings 
at large $N$ (the $1/N$ corrections).
We hope to report progress in this direction in the future.

\section{Beta deformation with mass term}
\label{RSMass}
A deformation of the matrix model with mass terms
and cubic couplings was introduced in \cite{RBBMN}.
This deformed model corresponds to membrane theory
on a pp-wave background with a constant four-form flux
\cite{RBBMNMembrane1}\cite{RBBMNMembrane2}.
General non-supersymmetric models are studied
from the membrane theory point of view
in \cite{RBIizuka}.
It is natural to try to simultaneously 
introduce the $\b$-deformation considered in this paper and the mass deformation.
This can be achieved, as we shall see below,
exploiting the linearity
of 
the equation of motion (\ref{RFEOMPPF})
for the three-form gauge field on the pp-wave background. 

We consider 
the pp-wave background with the
four-form flux which is a linear superposition 
of a constant part $f_{\a\b\c}$
and the linear (in transverse coordinates)  part $F^{(1)}$ 
for the $\beta$-deformation identified in
(\ref{RFBKGGeneralF1})(\ref{RFBKGGeneralF2}),
\begin{equation}
F_{+\a\b\c}=
f_{\a\b\c}
+
{F^{(1)}}_{+\a\b\c}.
\end{equation}
This solves the equation of motion for the gauge field
(\ref{RFEOMPPF}).

The equation of motion for the metric (\ref{RFEOMPPG})
becomes
\begin{eqnarray}
\frac{1}{2}\partial^\a \partial_\a G^{--}
&=&\frac{1}{12}
\Bigg(
f_{\a\b\c}f^{\a\b\c}
+
12
\sum_{I,J}
\left(
f_{\bI\bJ J}
F^{(1)}_{+ I J \bJ}
+
f_{I J \bJ}
F^{(1)}_{+ \bI \bJ J}
\right)
\non\\
&&
\qquad+
{F^{(1)}}_{+\a\b\c}
{{F^{(1)}}_+}^{\a\b\c}
\Bigg).
\end{eqnarray}
\Chk{-}
This equation can be solved by the ansatz
\begin{equation}
G^{--}=
\mu_{\a\b} x^\a x^\b
+
\k_{\a\b\c} x^\a x^\b x^\c
+
G^{--}_{(4)},
\end{equation}
where $G^{--}_{(4)}$ is the metric, quartic in coordinates,
given in (\ref{RFBKGGeneralG}).
The quadratic term, where $\mu_{\a\b}$ satisfies
\begin{equation}
{\mu_\alpha}^\alpha
=
\frac{1}{12} 
f_{\a\b\c}
f^{\a\b\c},
\label{RFEOMMu}
\end{equation}
is just the metric associated with the mass-deformation.
Finally, the cubic term, where $\k_{\a\b\c}$ satisfies
\begin{eqnarray}
\sum_J
2 (2\p)^2
\a^{(IJ)} f_{I J \bJ}
&=&3(\sum_J 2 \k_{I J \bJ}
+\k_{I 99}),
\label{RFEOMKappa1}
\\
\sum_J
2 (2\p)^2
\a^{(IJ)} f_{\bI\bJ J}
&=&3(\sum_J 2 \k_{\bI J \bJ}
+\k_{\bI 99}),
\label{RFEOMKappa2}
\end{eqnarray}
\Rder{
\begin{equation}
\sum_J
-2 (2\p)^2
\a^{(IJ)} f_{I J \bJ}
=3(\sum_J 2 \k_{I J \bJ}
+\k_{I 99})
\end{equation}
\begin{equation}
\sum_J
-2 (2\p)^2
\a^{(IJ)} f_{\bI\bJ J}
=3(\sum_J 2 \k_{\bI J \bJ}
+\k_{\bI 99})
\end{equation}
}
is the only essentially new term for the matrix model
which is simultaneously $\b$-deformed and mass-deformed.
Here we are using the notation
in which the nine real transverse coordinates 
are decomposed into four complex coordinates labelled by $I$,
its complex conjugates labelled by $\bI$,
and the one real direction $x^9$.

We see that there are ambiguities in $\k_{\a\b\c}$:
one can add
arbitrary traceless pieces to it.
Instead of considering the most general possibility,
we shall concentrate on a deformed model
having a more or less simpler form,
setting $\k_{I99}=0$, $\k_{\bI 99}=0$.
Later in this section, we will also exploit this ambiguity 
to construct a particularly tractable version of the
deformed matrix model.

The appearance of the cubic term might also be expected 
from the following consideration.
The constant part of the flux contributes the term,
\begin{equation}
\int \frac{1}{6} f_{\a\b\c} x^\c \{x^\a, x^\b\} d^2\s,
\label{RFPartHamiltonianContiConstflux}
\end{equation}
\Rder{
\begin{equation}
\int -\frac{1}{6} f_{\a\b\c} x^\c \{x^\a, x^\b\} d^2\s,
\end{equation}
}
\Chk{-}
to the Hamiltonian of the continuum theory. This term gives rise to the
term 
\begin{equation}
C_{\a\b\c} i \Tr X^\a[X^\b,X^\c] 
\label{RFCXXX}
\end{equation}
in the Hamiltonian of the matrix model. Here, $C_{\a\b\c}$
and $f_{\a\b\c}$ are related by
\begin{equation}
C_{\a\b\c}=
-\frac{1}{6} \frac{-P_-}{N} (2\pi T)^{-\frac{2}{3}}
 f_{\a\b\c},
\end{equation}
\Rder{
\begin{equation}
C_{\a\b\c}=
\frac{1}{6}
\frac{-P_-}{N} (2\pi T)^{-\frac{2}{3}}
f_{\a\b\c},
\end{equation}
}
which follows from (\ref{RFMRMultiplication})-(\ref{RFMRTr})
and the rescaling relation (\ref{RFRescalingX}).
Now, a natural guess about the deformation of the matrix model
in the present case
is to replace the commutators by the $*$-commutators
not only 
in the Yukawa and quartic scalar couplings,
but also in the cubic scalar couplings (\ref{RFCXXX}).
This means, in the continuum theory,
to replace the term 
(\ref{RFPartHamiltonianContiConstflux})
using the substitution rules 
(\ref{RFContinuumStarGeneral1})(\ref{RFContinuumStarGeneral2}).
We would then have cubic terms in the Hamiltonian.

However,
although the form of the cubic terms are correct, 
the over-all factor thus obtained turns out to be wrong 
(i. e. not consistent with the
equation of motion (\ref{RFEOMPPG})) by a factor of $\frac{2}{3}$.
One simple way (which is not so elegant) to 
resolve this issue is to introduce a new $*'$-product defined by
\begin{equation}
f*'g=e^{i \frac{3}{2} \pi \b(
Q^f_{(1)}
Q^g_{(2)}
-
Q^g_{(1)}
Q^f_{(2)})
} f g,
\end{equation}
and use it to deform the cubic couplings,
while using the original $*$-product for the quartic and Yukawa couplings.

Under this prescription, the deformed matrix model is given by,
\begin{eqnarray}
H&=&\Tr \Bigg( \frac{1}{2} (\P^\alpha)^2 
- \frac{1}{4} [X^\alpha, X^\beta]_*^2
+ \Psi^T \c^\a [X_\a,\Psi]_*
\non
\\
&&\quad 
+M_{\a\b}X^\a X^\b
+C_{\a\b\c} i X^\a[X^\b,X^\c]_{*'}
+
\kpm\frac{i}{4} C_{\a\b\c}
\Psi^T \c^{\c\b\a} \Psi
\Bigg),
\label{RFHamiltonianMassBetaMatrixModel}
\end{eqnarray}
where 
\begin{equation}
M_{\a\b}=\frac{1}{2}\left(\frac{-P_-}{N}\right)^2 (2\pi T)^{-\frac{4}{3}}
\mu_{\a\b}.
\end{equation}
It is easy to derive the fermionic term above,
since
the fermionic contribution 
to the continuum Hamiltonian is 
simply given by the sum of contributions 
for the purely mass-deformed case 
and for the purely $\b$-deformed case.
This can be easily seen from the argument in \ref{RSSFermionic},
in particular from (\ref{RFNorcorEVV2LC}).

It is known that, in the mass-deformed model, one finds
a stable solution corresponding to a spherical membrane
via the ansatz,
\begin{equation}
[X^\a, X^\b] \propto i {\e^{\a\b}}_{\c} X^\c, \qquad(\a,\b,\c=1,2,3),
\end{equation}
which corresponds to the ansatz,
\begin{equation}
\{x^\a, x^\b\} = \frac{a}{\tA}  {\e^{\a\b}}_{\c} x^\c, \qquad(\a,\b,\c=1,2,3),
\label{RFLBEpsilon}
\end{equation}
for the continuum theory.
Here, $\e^{\a\b\c}$ is the totally anti-symmetric tensor in three dimension;
it vanishes
if any of the indices take other values than
$1,2,3$.
The parameter $a$ is related to the flux $f_{\a\b\c}$; 
see (\ref{RFRelationfa}) below.
The appearance of $\tA$ in the above equation can be understood
by considering the rescaling of the $\s$-coordinates.

On the other hand,
in the $\b$-deformed model,
one finds stable solutions which correspond to 
membranes with the torus topology,
as has been shown in section \ref{RSSolution}.
It is therefore natural to expect that, for a model
having both parameter $\b$ and parameter $a$
for the mass-deformation, the stable solution
would have the topology of  a torus or a sphere,
depending on which of the two parameters is dominating. 
We note that a very similar phenomenon is studied 
in \cite{RBABHHS}. 
There, an one-parameter non-commutative algebra 
which interpolates between the 
non-commutative sphere and the non-commutative torus 
is constructed, together with
its explicit representations;
the information about the topology is
encoded in the eigenvalue distributions
in a manner proposed in \cite{RBMRTopology}.

In general, it seems difficult to study the stable solutions analytically.
However, we have found that by tuning the parameters
$\k_{I99}$, $\k_{\bI99}$ and $\mu_{99}$ one can obtain 
a particularly tractable class of the deformed models,
where one can explicitly demonstrate the interpolation
between a sphere and a torus.

The basic idea is to write a part of the Hamiltonian in the form,
\begin{equation}
\frac{\tA}{2(-P_-)}
\frac{T^2}{2}
\left(
\{x^\a,x^\b\}_*
- \frac{a}{\tA}
{\e^{\a\b}}_{\c} x^\c
\right)^2. 
\label{RFHamContiSquared}
\end{equation}
For simplicity we consider the simplest $\b$-deformation,
and we have denoted by $\{x^\a,x^\b\}_*$ the continuum counterpart of the
$*$-commutator, namely the right-hand side of 
(\ref{RFContinuumStar1})(\ref{RFContinuumStar2}).
As is well known, the advantage of writing the Hamiltonian 
in the above form is
that the zero-energy solution can be found
by solving the first order equation,
\begin{equation}
\{x^\a,x^\b\}_*
=\frac{a}{\tA} {\e^{\a\b}}_\c x^\c,
\label{RFDefLBEPsilon}
\end{equation}
which is a $\b$-deformed version of (\ref{RFLBEpsilon}).
From (\ref{RFHamContiSquared}), one can read off the background
$\mu_{\a\b}, f_{\a\b\c}, \k_{\a\b\c}$,
by comparing it to (\ref{RFHamiltonianContinuumFromBKG}). 
In particular, we have the relation between the parameter 
$a$ and $f_{\a\b\c}$,
\begin{equation}
f_{+\a\b\c}
=
- \frac{3 a T}{-P_-} \e_{\a\b\c},
\label{RFRelationfa}
\end{equation}
\Rder{
\begin{equation}
f_{+\a\b\c}^{(R)}
=
3 \frac{a T}{-P_-} \e_{\a\b\c},
\end{equation}
}
which also justifies the appearance
of $\tA$ in (\ref{RFLBEpsilon})(\ref{RFDefLBEPsilon}).
Unfortunately, the backgrounds thus read off do not satisfy
the equations of motion (\ref{RFEOMMu})-(\ref{RFEOMKappa2})
by themselves.  However, one can 
introduce extra backgrounds, $\k_{I99}$, $\k_{\bI 99}$, $\m_{99}$ 
such that the equations of motion are satisfied. 
The additional terms in the Hamiltonian introduced by this tuning
do not affect the stable solutions if we set $x^9=0$.

We have thus shown that there is a class of deformed matrix models
for which the equation (\ref{RFDefLBEPsilon}) gives the stable solutions (with $x^9$=0).
Using the complex coordinates
and setting unimportant scalars to zero,
(\ref{RFDefLBEPsilon})
becomes,
\begin{eqnarray}
&&
\{z,w\}_*
=
\frac{a}{\tA} \frac{i}{\sqrt{2}} z,
\label{RFDefLBEpsilonComplexZW}
\\
&&
\{z,\bw\}_*
=
\frac{a}{\tA} \frac{i}{\sqrt{2}} z,
\\
&&
\{z,\bz\}
=
\frac{a}{\tA} \left(-\frac{i}{\sqrt{2}}\right)
(w+\bw),
\\
&&\{w,\bw\}=0.
\label{RFDefLBEpsilonComplexWW}
\end{eqnarray}
with the help of
\begin{equation}
\epsilon_{z\bz w}= \frac{i}{\sqrt{2}},
\quad
\epsilon_{z\bz \bw}= \frac{i}{\sqrt{2}}.
\end{equation}

These equations reduce to those for the pure $\b$-deformation
if $a=0$, in which $z\sim e^{i \s^1}$, $w\sim e^{i \s^2}$ is a
solution (for $\b N=1$). If $\b$=0,
it reduce to those for the pure mass-deformation,
and then a sphere embedded in $x^1, x^2, x^3$
is a solution.
A natural ansatz for the general case, which interpolates between these two
solutions is,
\begin{eqnarray}
z&=&r(\s^2) e^{i\s^1},
\\
w&=&w(\s^2).
\end{eqnarray}
The equations (\ref{RFDefLBEpsilonComplexZW})-(\ref{RFDefLBEpsilonComplexWW})
then become ordinary differential equations,
\begin{eqnarray}
&&ir w'+
\b N\frac{(2\p)^2}{\tA} r w
= \frac{a}{\tA} \frac{i}{\sqrt{2}} r,
\label{RFVanishingDefLBODE1}
\\
&&
(r^2)'=-\frac{a}{\sqrt{2} \tA} (w+\bw),
\label{RFVanishingDefLBODE2}
\end{eqnarray}
where we have abbreviated $f'=\fracpp{f}{\s^2}$.
From (\ref{RFVanishingDefLBODE1}), we get,
\begin{equation}
w= C e^{i \b N \frac{(2\p)^2}{\tA} \s^2}
+ i \frac{a}{(2\p)^2 \sqrt{2} \b N}
\end{equation}
where $C$ is an integration constant. 
Substituting this to (\ref{RFVanishingDefLBODE2}),
we get 
\begin{equation}
r^2= D 
+ |C| \frac{\sqrt{2} a}{(2\p)^2\b N} \cos{\left(\b N\frac{(2\pi)^2}{\tA}\s^2\right)},
\end{equation}
where $D$ is another integration constant and 
we have chosen the phase of $C$ appropriately by shifting $\s^2$.

For $D> |C| \frac{\sqrt{2} a}{(2\p)^2\b N}$, the right-hand side is always positive.
One can take the range of $\s^2$ as $-\p< \s^2 <\pi$ without loss of generality.
Then $\tA$ equals to $(2\pi)^2$, and we see that 
$\b N$ should be an integer
because of the regularity of the solution.
The solution has the topology of $S^1\times S^1$, i. e. the torus.

For $D< |C| \frac{\sqrt{2} a}{(2\p)^2\b N}$, 
the right-hand side can become negative, whereas the left-hand side
is, by definition, always positive.
This implies that the range of $\s^2$ should be restricted to
$-l<\s^2<l$,
and at the point $\s^2=\pm l$ the radius $r$ should vanish.
Noting that $\tA=4 \p l$, we obtain,
\begin{equation}
0= D 
+ |C| \frac{\sqrt{2} a}{(2\p)^2\b N} \cos{\left(\b N \pi \right)}.
\label{RFCDSphere}
\end{equation}
In this case, the solution has the topology of a sphere.

To summarise, for the general case where $\b N$ is not an integer,
the solution has the topology of a sphere, and the integration 
constants $C$, $D$ are related by the equation (\ref{RFCDSphere}),
as a consequence of the boundary condition.
For the special case where $\b N$ is an integer, 
the solution has the topology of a torus, and
$C$, $D$ is only restricted by the inequality
\begin{equation}
D > |C| \frac{\sqrt{2} a}{(2\p)^2\b N}.
\end{equation}
Thus the dimension of the space of stable solutions enhances
at these special points.

One can also 
construct solutions to the corresponding matrix equations
\begin{equation}
[X^\a, X^\b]_*= i \frac{1}{N} T^{\frac{1}{3}}
(2\p)^{-\frac{2}{3}} a {\e^{\a\b}}_{\c} X^\c,
\end{equation}
by using the ansatz that $W$ is diagonal,
and only non-zero elements of $Z$ 
are those adjacent to diagonal elements.
The form of the matrix solutions is very similar to those given in
\cite{RBABHHS}.
Detailed formulae will be given elsewhere.


\section{Conclusion}
\label{RSConclusion}
In this paper, we have considered a class of deformation
for the matrix model of M-theory.
The form of the deformation, which consists in
modifying the Yukawa and quartic scalar couplings
by distributing flavour-dependent phase factors,
is motivated from a similar deformation
of $\mN=4$ SYM in four dimension.
In four dimension, this deformation has significance that
it preserves the conformal invariance of the
original $\mN=4$ SYM.
We have found that for the matrix model of M-theory,
the deformation is also special in that
it admits M-theory interpretation: 
the deformed model can be considered as the matrix model
of M-theory on a certain curved background,
since it is equivalent to a regularised version of 
supermembrane theory
on that background.

It is remarkable that the deformation, introduced 
from a rather mathematical analogy to four-dimensional field theory,
has a natural eleven-dimensional interpretation.
Indeed, this interpretation requires strong consistency,
as the identified background should satisfy 
the supergravity equations of motion.
We have verified that they are indeed satisfied including an over-all factor,
without any artificial tuning of parameters. 
One might say that,
somehow, the $\beta$-deformation ``knows'' the eleven-dimensional
supergravity.
It is hard to believe that this high degree of consistency 
is a mere coincidence.
It would be fascinating if one could find
a framework to understand this consistency in a natural fashion. 

In general, pp-wave backgrounds
arise as a result of a limiting procedure called the Penrose limit;
in order to obtain a physical interpretation of our background,
it might be useful to consider what backgrounds would 
reduce to our pp-wave background under the Penrose limit. 

The deformed model also seems to contain
interesting physics.
It has stable solutions, which corresponds to 
toric membranes with the simple $S^1\times S^1$ shape,
for some particular values of the deformation parameter. 
The solution has classical flat directions, which 
correspond to the radii of two circles.
To consider quantum corrections to these flat directions
is an interesting problem.
Also, we have found that some configurations
which are physicaly distinct in conventional membrane theory
should be considered as the same object in the matrix model.
We have argued that this might be the reflection of the non-Abelian nature of
membranes.
We have also studied the $\b$-deformed model with the mass terms
and found that, for a particular class of models,
there are stable membrane configurations which interpolates between a membrane 
with the topology of a torus and a membrane with the topology of a sphere.

Finally,
we wish to raise a few directions
one might pursue concerning our deformed matrix model.
It will be interesting to relate the model to 
ten-dimensional type IIA string theory,
by compactifying the $x^9$ direction, which  is not touched 
even for the most general deformation. 
Alternatively, one might
compactify the $x^+$ direction;
the stable solutions discussed in section \ref{RSSolution}
will be wrapped in the $x^+$ direction, and hence can be considered
as a string worldsheet with the topology of a torus. 
This might allow one to
interpret the stable solutions as
saddle points of the path integral of 
suitably Euclideanised 
type IIA string theory on a curved background.

The pp-wave background considered in this paper
is with the metric and the four-form flux
respectively given by
quartic and linear polynomials of the transverse coordinates.
It would be interesting to consider the generalisation of
our matrix model which 
corresponds to a similar pp-wave metric  
with more general higher-order polynomial.

One can ask many questions about this model,
other than those already mentioned, such as the 
classification of BPS states, scattering of various objects,
in particular the gravitons.
We hope that this model would serve as a good place to further explore 
the physics of membranes and the matrix model.

I would like to thank S. Theisen, for continuous encouragement,
discussion and useful comments.
I would also like to thank
J. Arnlind,
S. Ananth, 
K. Hashimoto, 
J. Hoppe,
S. Iso,
N. Kim, 
M. Kato,
Y. Kazama, 
T. Kuroki,
S. Kovacs,
Y. Okawa,
J. Plefka,
T. Shimada,
S. Sugimoto, 
F. Sugino,
T. Yoneya, 
K. Yoshida,
for encouragement, discussion, and useful comments.
The author is grateful for the grant SFB647 ``Raum-Zeit-Materie'',
and is very thankful to his colleagues in the Albert-Einstein-Institut.
Most of this work is first presented at the school ``Aspects of Membrane Dynamics''
in KTH, Stockholm, on June 11-23, 2007.

\appendix
\renewcommand{\theequation}{\Alph{section}.\arabic{equation}}
\makeatletter
\@addtoreset{equation}{section}

\section{Notations and Conventions}
\label{RSANotation}
In this appendix we summarise  notations and conventions
used in this paper. See also appendix \ref{RSASuper} for
some of the notations and conventions which are specific to 
subsection \ref{RSSFermionic}.

Our signature of the metric is
\begin{equation}
\eta_{\mu\nu} = \mbox{diag} (- + \cdots +).
\end{equation}

The meaning and the range of various indices are as follows,
\begin{eqnarray*}
\m,\n,\cdots &:& \mbox{vector indices for the spacetime; run through }0,\cdots, 10,
\\
 \a,\b,\cdots&:& \mbox{indices for transverse directions; run through }1,\cdots,9,
\\
a,b,\cdots &:& \mbox{spinor indices;
usually run through $1,\cdots,32$ in subsection \ref{RSSFermionic}.}
\\
&&\mbox{run through $1, \cdots,16$ when explicitly stated,}
\\
i,j,\cdots&:& \mbox{either $U(N)$ matrix indices which run through $1,\cdots,N$,}
\\
&& \mbox{or the worldvolume vector indices which run through $0,1,2$,}
\\
r,s,\cdots&:& \mbox{parametrise the worldvolume
spacelike coordinates; run through $1,2$,}
\\
A, B,\cdots&:&
\mbox{only used in subsection \ref{RSSFermionic};}
\\
&&
\mbox{collectively denotes $\mu, \nu,\cdots $ indices and
$a,b,\cdots$ indices;\ } A=(\mu,a).
\end{eqnarray*}
In subsection \ref{RSSFermionic},
we distinguish 
the tangent space indices
$\hA,\hB,\hmu,\hnu,\ha,\hb,\hat{+},\hat{-},\cdots$ from the curved space indices
$A,B,\m,\n,a,b,+,-,\cdots$.

We use the $16\times16$ real and symmetric $SO(9)$ gamma matrices,
\begin{equation}
\c^\a \c^\b +\c^\b \c^\a =2 \delta^{\a\b}.
\end{equation}
We use
\begin{equation}
\gamma^{\a_1 \cdots \a_n}
=
\frac{1}{n!}
\left(\c^{\a_1}\cdots
\c^{\a_n} \pm \mbox{($n!-1$ permutations)}
\right).
\end{equation}

We define the total area of the base space of a 
$x^+$-slice of membranes, parametrised by $\s^1$, $\s^2$,
as
\begin{equation}
\tA=\int d^2\s.
\end{equation}

%

Our lightcone conventions are,
\begin{eqnarray}
&&v^{\pm}=\frac{v^0\pm v^{10}}{\sqrt{2}},
\\
&&\eta_{+-}= -1,\quad \eta_{-+}= -1,
\\
&&w_{\pm}= -w^{\mp}
=-\frac{w^0\mp w^{10}}{\sqrt{2}}.
\end{eqnarray}

\XXX{Anything else?}

\section{Details about rescaling}
\label{RSARescaling}
We  describe here
the rescaling of dynamical variables and the time coordinate,
necessary to bring the matrix-regularised supermembrane theory 
(on flat spacetime) into the normalised matrix model form
(\ref{RFHamiltonianOriginalMatrixModel})(\ref{RFConstraintUN}).
This rescaling is not affected by the introduction of
the deformation.

The Dirac brackets (or the Poisson brackets) 
of continuum membrane theory
are given by, for bosonic variables, 
\begin{equation}
\{x^\a(\s'),\mP^\b(\s'')\}_{\mbox{D.B.}}
=
\d^{\a\b} \d^2(\s'-\s'').
\label{RFDBContiXP}
\end{equation}
The Dirac brackets 
between 
the matrices corresponding to $x$ and $\mP$,
$\rho(x)=\hat{x}$ and $\rho(\mP)=\hat{\mP}$, 
are given by, 
\begin{equation}
\{
\left(\hat{x}^\a\right)^i_j
,
\left(\hat{\mP}^\b\right)^k_l
\}_{\mbox{D.B.}}
=\frac{N}{\tA}
\d^{\a\b}\d^i_l \d^k_j,
\quad (i,j,k,l=1,\cdots N).
\label{RFDBBosonicBeforeR}
\end{equation}
The factor $\frac{N}{\tA}$ arises because
(i) 
integration over $\s''$ in (\ref{RFDBContiXP}) corresponds
to taking partial trace in (\ref{RFDBBosonicBeforeR}), 
say contracting indices $i$ and $j$, multiplied
by a factor $\frac{\tA}{N}$,
because of (\ref{RFMRTr}),
and (ii)
the function on $(\s_1,\s_2)$-space taking the constant value $1$
corresponds to identity matrix, as can be seen from (\ref{RFMRMultiplication}).
Alternatively, one can derive (\ref{RFDBBosonicBeforeR}) from the general
relation between 
the Hamilton formalism
and 
the variational principle
on the phase space $\delta\int \left(p\dot{q}-H(q,p)\right) dt =0$, 
with the help of the relation
$\int \mP \dot{x} d^2 \s \approx
\frac{\tA}{N} \Tr \hat{\mP} \dot{\hat{x}}$.

Similarly, Dirac brackets for the fermionic variables in
the continuum theory (\ref{RFDBFermionMembrane}), 
\[
\sqrt{2} \frac{(-P_-)}{\tA}
\{
\psi^a(\s')
,
\psi^b(\s'')
\}_{\mbox{D.B.}}
=
-\frac{i}{2} \d^{ab} \d^2(\s'-\s''),
\]
become
\begin{equation}
\sqrt{2} \frac{(-P_-)}{N}
\{
\left(\hat{\psi}^a\right)^i_j
,
\left(\hat{\psi}^b\right)^k_l
\}_{\mbox{D.B.}}
=
-\frac{i}{2}\d^{ab}\d^i_l \d^k_j,
\label{RFDBFermionicBeforeR}
\end{equation}
after regularisation, where $a,b=1,\cdots,16$.

By applying (\ref{RFMRMultiplication})-(\ref{RFMRTr})
to the continuum Hamiltonian,
\begin{equation}
H
=
\int
\left(
\frac{\tA}{2(-P_-)} 
\left(
\left(\mP^\alpha\right)^2
+
\frac{1}{2} \{x^\a, x^\b\}^2
\right)
+
\kpm \sqrt{2}i
\psi^T \c^\a \{x^\a,\psi\}
\right)
d^2\s,
\label{RFHamiltonianFlatSuper}
\end{equation}
we get the Hamiltonian for regularised theory,
\begin{eqnarray}
H&=&
\frac{\tA}{N}
\Tr
\Bigg(
\frac{\tA}{2(-P_-)}
\left(
\left(\hat{\mP}^\a\right)^2
+
\frac{1}{2}
\left(
\frac{2 \pi N}{i\tA}
[\hx^\a, \hx^\b]^2
\right)^2
\right)
\non\\
&&
+
\kpm i \sqrt{2} 
\hat{\psi}^T
\c^\a
\left(\frac{2\p N}{i \tA}
[\hx^\a,\hat{\psi}]
\right)
\Bigg).
\label{RFHamBeforeR}
\end{eqnarray}
Similarly
the constraint (\ref{RFConstraintAPDSuper}),
\[
0=
\{
\mP^\a, x^\a
\}
+
\frac{-P_-}{\tA}
i \sqrt{2}
\{
\psi^T, \psi
\},
\]
becomes
\begin{equation}
0=\frac{2\p N}{i\tA}
[\hat{\mP}^\a, \hat{x}^\a]
+
\frac{-P_-}{\tA} i \sqrt{2}
\frac{2\p N}{i\tA}
2\hat{\psi}^T\hat{\psi}.
\label{RFConstrBeforeR}
\end{equation}

These relations 
(\ref{RFHamBeforeR})(\ref{RFConstrBeforeR})(\ref{RFDBBosonicBeforeR})(\ref{RFDBFermionicBeforeR})
can be brought into the form 
(\ref{RFHamiltonianOriginalMatrixModel})-(\ref{RFDBOriginalMMFermion})
by using the rescaling
\begin{eqnarray}
&&
d\t
=
\frac{-P_-}{N}
(2\pi)^{-\frac{2}{3}}T^{-\frac{2}{3}} dt,
\\
&&
{\hat{x}}^\a
=
(2\p)^{-\frac{1}{3}} T^{-\frac{1}{3}} X^\a,
\label{RFRescalingX}
\\
&&
{\hat{\mP}}^\a=
\frac{N}{\tA}(2\p)^{\frac{1}{3}}
T^{\frac{1}{3}}
\Pi^\a,
\\
&&
\Psi=
2^{\frac{1}{4}}
\sqrt{\frac{-P_-}{N}} \hat{\psi},
\end{eqnarray}
where we have restored the membrane tension $T$.

\section{Details for the fermionic part}
\label{RSASuper}
We compile in this appendix some of the detailed
formulae and conventions for supermembrane theory, used in subsection
\ref{RSSFermionic}.

We use the following anti-commutation relation 
for the $32\times32$ $SO(1,10)$ gamma matrices,
\begin{equation}
\C^\hmu \C^\hnu +\C^\hnu \C^\hmu =2 \eta^{\hmu\hnu}.
\label{RFGammaAC32}
\end{equation}
We use the representation 
in which all gamma matrices are real,
and $\C^{\hat{0}}$ is anti-hermitian and other gamma matrices are hermitian.
More explicitly, we use
\begin{eqnarray}
\C^{\hat{0}}&=&
\left(
\begin{array}{cc}
0&1\\
-1&0
\end{array}
\right)
\tensorp 1_{16},
\\
\C^\halpha&=&
\left(
\begin{array}{cc}
1&0\\
0&-1
\end{array}
\right)
\tensorp \c^\alpha,
\\
\C^{\hat{10}}&=&
\left(
\begin{array}{cc}
0&1\\
1&0
\end{array}
\right)
\tensorp 1_{16},
\end{eqnarray}
where $\gamma^\a$ are the 
$16\times 16$ real and symmetric SO(9) gamma matrices, and
$1_{16}$ denotes the $16\times 16$ unit matrix.
For lightcone directions, we have,
\begin{eqnarray}
\C^{\hat{+}}&=&
\left(
\begin{array}{cc}
0&\sqrt{2}\\
0&0
\end{array}
\right)
\tensorp 1_{16},
\\
\C^{\hat{-}}&=&
\left(
\begin{array}{cc}
0&0\\
-\sqrt{2}&0
\end{array}
\right)
\tensorp 1_{16},
\label{RFGammaMinus}
\end{eqnarray}
so that, under the lightcone gauge condition,
$\C^+ \th = 0$, 32-component spinor $\th$ reduces to
the 16-component SO(9) spinor $\psi$ as,
\begin{equation}
\th= 
\left(
\begin{array}{c}
\psi \\
0
\end{array}
\right).
\label{RFDef16Spinor}
\end{equation}
The gamma matrices defined above have one upper and one lower indices.
We use gamma matrices with two lower indices,
\begin{equation}
{\C^{\hmu_1\cdots\hmu_n}}_{\ha\hb} = \mC_{\ha\hc} 
{{\C^{\hmu_1\cdots\hmu_n}}^{\hc}}_{\hb},
\label{RFGammaLL}
\end{equation}
where $\mC$ is the charge conjugation matrix, which is proportional
to $\C^{\hat{0}}$ in our representation.
We choose the convention, $\mC=\C^{\hat{0}}$.

The transformation law of the $\kappa$-symmetry is given by
the following ansatz,~\footnote{
We could introduce a minus sign in (\ref{RFKappaS}).
It can be absorbed by flipping the orientation
on the world-volume for a particular configuration
of membranes, without any change of the theory in total.
}
\begin{eqnarray}
&& 
\delta x^A {E_A}^\hmu=0,
\label{RFKappaV}
\\&&
{\Gamma^\ha}_\hb
\delta x^A {E_A}^\hb
=
\kpm \delta x^A {E_A}^\ha,
\label{RFKappaS}
\end{eqnarray}
where $\Gamma^2=1$ is defined by
\begin{equation}
{\C^\ha}_\hb
 = \sqrt{-\det{h_{ij}}} 
\pi^{0\hmu}
\pi^{1\hnu}
\pi^{2\hrho}
{(\C_{\hmu\hnu\hrho})^\ha}_\hb,
\qquad 
\pi^{i\mu} = 
h^{ij} {\pi_j}^\hmu.
\end{equation}

The covariant derivative only acts on the tangent
indices and our convention is such that,
\begin{equation}
D_A v^\hB
=
v^\hB
-
{\O_{A\hC}}^\hB v^\hC
(-1)^{\hC\hB}.
\end{equation}
\Rder{
}
Here, indices in the exponent of $(-1)$
are to be substituted by 
$0$
if they are bosonic 
and by $1$
if they are fermionic. 
These factors are 
necessary in order 
to maintain the correct transformation
property of superspace tensors.
In superspace formulation of supergravity,
the gauge symmetry acting on the tangent space 
is taken to be the local Lorentz symmetry.
As a consequence, the connection satisfies,
\begin{eqnarray}
{\O_{A\hb}}^\ha
=
\frac{1}{4}
{\left(\C^{\hmu\hnu}\right)^\ha}_\hb
\O_{A\hnu\hmu},
\\
{\O_{A\hb}}^\hmu=0, \quad
{\O_{A\hmu}}^\hb=0.
\end{eqnarray}
\Rder{
\begin{equation}
{{\O^\ha}_\hb}_A=
\frac{1}{4} \C^{\mu\nu} \O_{\hmu\hnu A}
\end{equation}
\begin{equation}
{{\O^\hmu}_\hb}_A=0
\end{equation}
\begin{equation}
{{\O^\ha}_\hmu}_A=0.
\end{equation}
}
\Chk{*}
The derivatives by anti-commuting variables are usual left derivatives.
\Rder{
unusual right derivatives
}

The torsion tensor 
${T_{AB}}^\hC$ is defined by,
\begin{equation}
{T_{AB}}^\hC
=
\sum_{AB}
D_A {E_B}^\hC
=
\sum_{AB}
\left(
\partial_A {E_B}^\hC
-
{\O_{A\hD}}^\hC{E_B}^\hD
(-1)^{(\hD+\hC)(B+\hD)}
\right),
\label{RFDefTorsion}
\end{equation}
where $\sum_{AB}$ stands for the graded anti-symmetric summation
over all independent permutations of indices $A$ and $B$.
We transform two lower indices of the torsion tensor into tangent space indices 
via the relation,
\begin{equation}
{T_{AB}}^\hC
=
{E_B}^\hE
{E_A}^\hD
{T_{\hD \hE}}^\hC
(-1)^{A(B+\hE)}.
\label{RFDefTorsionHat}
\end{equation}
Similarly, the four-form field strength tensor
is defined, together with their
tangent space components, as
\begin{eqnarray}
&&F_{A_1 A_2 A_3 A_4}
=
\sum_{A_1 A_2 A_3 A_4}
\partial_{A_1} A_{A_2 A_3 A_4},
\label{RFDefSuperF}
\\
&&F_{A_1A_2A_3A_4}
=
{E_{A_4}}^{\hB_4}
{E_{A_3}}^{\hB_3}
{E_{A_2}}^{\hB_2}
{E_{A_1}}^{\hB_1}
F_{\hB_1\hB_2\hB_3\hB_4}
\non\\
&\times&
(-1)^{
(A_2+\hB_2)A_1
+(A_3+\hB_3)(A_1+A_2)
+(A_4+\hB_4)(A_1+A_2+A_3)
}. 
\label{RFDefSuperFHat}
\end{eqnarray}
\Rder{
\begin{equation}
F_{\hA \hB \hC \hD}
=A_{\hA \hB \hC, \hD} + perm.
\end{equation}
\begin{equation}
F_{\hA_1\hA_2\hA_3\hA_4}
{E^{\hA_4}}_{B_4}
{E^{\hA_3}}_{B_3}
{E^{\hA_2}}_{B_2}
{E^{\hA_1}}_{B_1}
(-1)^{
B_4(\hA_3+B_3)
+(B_3+B_4)(\hA_2+B_2)
+(B_2+B_3+B_4)(\hA_1+B_1)
}
=
F_{B_1B_2B_3B_4}
\end{equation}
}

The torsion and the field strength satisfy
the Bianchi identities, as a result of
the (anti-) commutativity of
the differential operators.
Defining the curvature by,
\begin{eqnarray}
&&{R_{AB\hC}}^\hD
=- \sum_{AB}
\left(\partial_A {\O_{B\hC}}^\hD
+ {\O_{A\hC}}^\hE{\O_{B\hE}}^\hD
(-1)^{\hE B}
\right),
\label{RFDefSuperCurvature}
\\
&&{R_{AB\hC}}^\hD
=
{E_B}^\hF
{E_A}^\hE
{R_{\hE \hF \hC}}^\hD
(-1)^{(B+\hF)A},
\end{eqnarray}
the Bianchi identities read,
\begin{eqnarray}
&&\sum_{\hA\hB\hC}
-D_\hA {T_{\hB\hC}}^\hD
-
{T_{\hA\hB}}^\hE
{T_{\hE\hC}}^\hD
=
\sum_{\hA\hB\hC}
-{R_{\hA\hB\hC}}^\hD\ ,
\label{RFBianchiDT}
\\
&&\sum_{\hA\hB\hC\hD\hE}
\left(
D_\hA F_{\hB\hC\hD\hE}
+
{T_{\hA\hB}}^\hF
F_{\hF\hC\hD\hE}
\right)
=0.
\label{RFBianchiDF}
\end{eqnarray}
\Rder{
\begin{equation}
\sum_{\hB\hC\hD}
-({T^\hA}_{\hB\hC})_{;\hD}
-
{T^\hA}_{\hB\hE}
{T^\hE}_{\hC\hD}
=
\sum_{\hB\hC\hD}
+{R^\hA}_{\hB\hC\hD}
\end{equation}
\begin{equation}
\sum_{\hA\hB\hC\hD\hE}
\left(
F_{\hA\hB\hC\hD;\hE}
+F_{\hA\hB\hC\hF} {T^\hF}_{\hD\hE}
\right)
=0
\end{equation}
}

By applying the Bianchi identities
to the superspace constraints 
(\ref{RFSuperspaceConstraintsT})-(\ref{RFSuperspaceConstraintsZero}),
one obtains many relations between components of 
the torsion, the  curvature, and the field strength.
For example, one can show that ${T_{\ha \hb}}^\hc$ vanishes.
Of particular importance is the relation,
\begin{equation}
{T_{\hmu\hb}}^\ha
= 
\kmp \frac{1}{36}
{\left(F_{\hmu\hsigma_1\hsigma_2\hsigma_3} \C^{\hsigma_3 \hsigma_2 \hsigma_1}
+\frac{1}{8} F_{\hsigma_1\hsigma_2\hsigma_3\hsigma_4}
{\C^{\hsigma_4 \hsigma_3 \hsigma_2 \hsigma_1}}_\hmu\right)^\ha}_\hb.
\end{equation}
\Rder{
\begin{equation}
{T^\ha}_{\hb \hmu} 
= 
-\frac{1}{36} \frac{s_1}{s_2} 
{\left(F_{\hmu\hsigma_1\hsigma_2\hsigma_3} \C^{\hsigma_3 \hsigma_2 \hsigma_1}
+\frac{1}{8} F_{\hsigma_1\hsigma_2\hsigma_3\hsigma_4}
{\C^{\hsigma_4 \hsigma_3 \hsigma_2 \hsigma_1}}_\hmu\right)^\ha}_\hb
\end{equation}
}
It is easy to check the numerical coefficients in the above expression,
using vector-vector-vector-spinor-spinor components of 
(\ref{RFBianchiDF}).

The gauge symmetry acting on the background fields, namely,
the general coordinate invariance and the local Lorentz transformation,
and the gauge transformation of the three form gauge fields, 
are fixed by the conditions
\begin{eqnarray}
\theta^b \th^a (\partial_a {E_b}^\hA - \partial_b{E_a}^\hA)&=&0,
\label{RFNorcorGC}
\\
\theta^a \O_{a \hmu\hnu} &=&0,
\label{RFNorcorLocalLorentz}
\\
\th^a A_{a AB}&=&0,
\label{RFNorcor3form}
\end{eqnarray}
\Rder{
\begin{eqnarray}
\O_{\hmu\hnu a} \theta^a &=&0 \\
({E^\hA}_{a,b} - {E^\hA}_{b,a}) \theta^b \th^a &=&0 \\
A_{ABa}\th^a&=&0\\
\end{eqnarray}
}
respectively.

Apart from this, one in general set,
\begin{eqnarray}
\left. {E_b}^\ha\right|_{\th=0}&=&{\delta^\ha}_b \\
\left. {E_a}^\hmu\right|_{\th=0}&=& 0
\end{eqnarray}
\Rder{
\begin{eqnarray}
\left. {{E^\ha}_b}\right|_{\th=0}&=&{\delta^\ha}_b, \\
\left. {{E^\hmu}_a}\right|_{\th=0}&=& 0,
\end{eqnarray}
}
by using the gauge symmetries. This makes the
correspondence between 
the $\th=0$ part of the superfields 
and the component supergravity fields simple.

We shall briefly outline the derivation
of the expression for the $(\th)^2$ part of the vector-vector component
of the supervielbein (\ref{RFNorcorEVV2}).
The condition (\ref{RFNorcorLocalLorentz}) implies
\begin{equation}
\O_{a \hmu\hnu}|_{\th=0} =0,
\quad
\left.\left(
\partial_b\O_{a \hmu\hnu}
-
\partial_a\O_{b \hmu\hnu}
\right)
\right|_{\th=0} =0,
\quad\cdots
\end{equation}
Using these relations and the definition of the torsion
(\ref{RFDefTorsion}), we obtain
\begin{equation}
\frac{1}{2} \th^b \th^a 
\left.\left(\partial_a \partial_b {E_\n}^\hmu\right)\right|_{\th=0}
=
\frac{1}{2} \th^b \th^a 
\left.\left(
\partial_a
\left(
{T_{b\n}}^\hmu
+
\partial_\nu {E_b}^\hmu
\right)
\right)\right|_{\th=0}.
\label{RFNorcorEVV2Step1}
\end{equation}
One can show that the following contribution from the first term in the 
brackets above
is the only non-vanishing term under the superspace constraints
(\ref{RFSuperspaceConstraintsT})-(\ref{RFSuperspaceConstraintsZero}),
\begin{equation}
-\frac{1}{2} \th^b \th^a 
\left.\left(
\left(\partial_a 
{E_\nu}^\hd\right) {E_b}^\hc
{T_{\hc \hd}}^\hmu\right)\right|_{\th=0}.
\end{equation}
By manipulating $\partial_a {E_\nu}^\hd |_{\th=0}$
in a similar manner to the manipulation in (\ref{RFNorcorEVV2Step1}),
we obtain (\ref{RFNorcorEVV2}).

The starting point to construct the lightcone gauge formalism
for supermembrane theory is the set of phase space constraints.
We denote the canonical momenta of $x^\mu$ and $\th^a$
by $\mP_\mu$ and $\mP_a$. It is convenient to define $\tilde{\mP}$'s,
which are the contributions to the momenta from the $S_1$ in the action
(\ref{RFActionMembraneCurvedSuperTotal})-(\ref{RFActionMembraneCurvedSuperGauge}),
by
\begin{eqnarray}
&&
\tilde{\mP}_\mu=
\mP_\mu
+\frac{1}{2}\{x^C, x^B\}A_{\mu BC},
\\
&&
\tilde{\mP}_a=
\mP_a
-\frac{1}{2}\{x^C, x^B\}A_{BCa}.
\end{eqnarray}
\Rder{
\begin{eqnarray}
&&\tilde{\mP}_A=
\mP_A
-
A_{ABC}^R(-\frac{1}{2})\{x^C, x^B\}
\end{eqnarray}
}
In terms of them, 
the phase space constraints
are expressed as,
\begin{eqnarray}
&&
\tilde{\mP}_\mu
\tilde{\mP}_\nu
G^{\m\n}
+
(h_{11}h_{22} -h_{12}h_{21})=0,
\label{RFConstraintTauSuper}
\\
&&
\tilde{\mP}_\m \partial_r x^\m
+
\tilde{\mP}_a \partial_r x^a
=0,
\label{RFConstraintSigmaSuper}
\\
&&\tilde{\mP}_a
=
-{E_a}^\hnu {e_\hmu}^\mu \tilde{\mP}_\mu,
\label{RFConstraintFermionicMomenta}
\end{eqnarray}
where ${e_\hnu}^\mu$ is defined by,
\begin{equation}
{e_\hnu}^\mu {E_\mu}^\hrho
={\delta_\hnu}^\hrho.
\end{equation}
\Rder{
\begin{eqnarray}
&&
\tilde{\mP}_A x^A_{,r}=0
\\
&&
\tilde{\mP}_\mu
\tilde{\mP}_\nu
G^{\m\n}
+
(h_{11}h_{22} -h_{12}h_{21})=0
\end{eqnarray}
\begin{equation}
\tilde{\mP}_a
=
\tilde{\mP}_\n {e^\n}_\hmu {E^\hmu}_a
\end{equation}
\begin{equation}
{E^\hmu}_\nu {e^\nu}_\rho ={\delta^\hmu}_\rho
\end{equation}
}


\end{document}